\documentclass{jfm}
\usepackage{amsmath}
\usepackage{graphicx}
\usepackage{epstopdf, epsfig}
\usepackage{color}

\newcommand{\lra}[1]{\langle #1 \rangle }
\newcommand{\mc}[1]{\mathcal{#1}}
\newcommand{\mb}[1]{\mathbf{#1}}
\newcommand{\bds}[1]{\boldsymbol{#1}}
\newcommand{\bfnabla}{\boldsymbol{\nabla}}

\DeclareFontFamily{OT1}{pzc}{}
\DeclareFontShape{OT1}{pzc}{m}{it}{<-> s * [1.10] pzcmi7t}{}
\DeclareMathAlphabet{\mathpzc}{OT1}{pzc}{m}{it}

\shorttitle{A Lagrangian pdf model for collisional turbulent fluid--particle flows}
\shortauthor{A.\ Innocenti, R.~O.\ Fox, and S.\ Chibbaro}

\title{A Lagrangian probability-density-function model for collisional turbulent fluid--particle flows. I. Model derivation}

\author{A.\ Innocenti\aff{1,2},
   R.~O.\ Fox\aff{3},  \and S.\ Chibbaro\aff{1} \corresp{\email{chibbaro@ida.upmc.fr}}}

\affiliation{\aff{1}Sorbonne Universit\'e, CNRS, Institut Jean Le Rond d'Alembert, F-75005 Paris, France
\aff{2}Dipartimento di Ingegneria Civile e Industriale, Universit\`a di Pisa, Via G. Caruso 8, 56122 Pisa, Italia
\aff{3}Department of Chemical and Biological Engineering, 618 Bissell Road, Iowa State University, Ames, IA 50011-1098, USA}

\begin{document}

\maketitle
\begin{abstract}
Inertial particles in turbulent flows are characterised by preferential concentration and segregation and, at sufficient mass loading, dense particle clusters may spontaneously arise due to momentum coupling between the phases. These clusters, in turn, can generate and sustain turbulence in the fluid phase, which we refer to as cluster-induced turbulence. 
In the present theoretical work, we tackle the problem of developing a framework for the stochastic modelling of moderately dense particle-laden flows, based on a Lagrangian formalism, which naturally includes the Eulerian one. A rigorous formalism and a general model have been put forward focusing, in particular, on the two ingredients that are key in moderately dense flows, namely,
two-way coupling in the carrier phase, and the decomposition of the particle-phase velocity into its spatially correlated and uncorrelated components.
Specifically, this last contribution allows to identify in the stochastic model the contributions due to the correlated fluctuating energy and to the granular temperature of the particle phase, which determines the time scale for particle--particle collisions. 
Applications of the Lagrangian probability-density-function model developed in this work to moderately dense particle-laden flows are discussed in a companion paper. 
\end{abstract}

\begin{keywords}
particle-laden flow, multiphase turbulence, Lagrangian pdf model, turbulence modulation
\end{keywords}
\section{Introduction} \label{sec:introduction}

The study of the dispersion of solid particles in a fluid has received much attention since the pioneering work on Brownian motion \citep{einstein1905theory,von1906kinetischen}. This type of flow has great impact for many engineering and environmental problems, but yet remains poorly understood, especially in the dense regime \citep{stickel2005fluid,forterre2008flows}. Even more underdeveloped is the situation concerning the turbulent regime and its modelling, which has received systematic attention only very recently \citep{Fox2014}. 
In general, if fine particles are dispersed in a Newtonian liquid, one speaks of suspensions \citep{guazzelli2011physical}, and when the suspension is dry, that is the fluid is gaseous, one has a granular material \citep{kadanoff1999built,Jae_96}. 
Even if the impact of the liquid phase, which has a mass density comparable to that of particles, on the dispersed phase is more important than the gaseous one, it has been shown that in many respects suspensions and granular flows have similar properties \citep{boyer2011unifying}.

It is more difficult to distinguish precisely the different regimes based on the volume fraction of particles dispersed in the flow \citep{elghobashi1992direct}.
When the particle volume fraction is very small, in the dilute regime, the fluid phase carries the particles, there is negligible momentum exchange between particles and no feedback of these on the fluid. Dilute particle-laden flows are encountered in nature very frequently and most of the time they are turbulent, and this regime has been vastly studied in the past \citep{balachandar2010turbulent}. Of course, it is not possible for the particles to induce large-scale turbulent motion in the fluid phase for this case. 

The back reaction exerted by the particles on the fluid gives rise to an extra complexity in modelling \citep{elghobashi1994predicting}. The understanding of the mechanisms at play, even at a purely qualitative level, is an important subject of research for developing efficient models of relevance to applications \citep{Gualtierietal2017}. For example, an instance where turbulence modulation plays a key role is in the process of planet formation \citep{johansen2007rapid}. Many questions on the modelling of turbulence modulation by particles remain also open in engineering applications. These include turbulent sprays \citep{jenny2012modeling} or fuel droplets in combustion chambers \citep{post2002modeling}, where two-way coupling is expected to enhance heat transfer and macroscopic chemical reaction rates. 

Here we consider moderately dense turbulent flows, in which the mass fraction is sufficient to trigger a feedback of the particle phase on the fluid, that is a two-way coupling between phases is present. When the particle volume fraction is high enough, collisions will be also at play. 
When the flow is dense, a hydrodynamic approach to the particle phase appears physically reasonable, thanks to a separation of scales, and this phase represents a compressible fluid. In such a situation, it has been shown that the particle phase can display turbulence and via the exchange between phases can even induce turbulence in an initially laminar fluid phase, because of  particle clustering \citep{capecelatro2014numerical,Fox2015}. 

From a modelling point of view, while in the turbulent dilute regime Lagrangian approaches have been shown to be superior \citep{Min_01,Min_04,Pei_06}, a two-fluid approach is generally used to handle dense flows, which seems physically sound on the basis of the considerations above \citep{crowe2011multiphase}. 
Unfortunately, the turbulence models developed for dense particle-laden flows in analogy with the single-phase ones \citep{Pope_turbulent}, have lacked a rigorous foundation and are often affected by flaws \citep{simonin1996continuum,peirano1998fundamentals}. Only recently, a two-fluid approach rigorously derived from an underlying  kinetic model has been presented \citep{Fox2014}, and it turned out key to distinguish between the particle-phase granular energy and turbulent kinetic 
energy, which can be rephrased as the spatially uncorrelated and correlated components of the fluctuating particle velocity fields, as originally introduced by \cite{fevrier2005partitioning} in the case of dilute flows.

In this work, we develop a Lagrangian approach to moderately dense turbulent flows. 
In particular, we propose a two-way coupled model
in a form that clearly separates the correlated and uncorrelated components.
There were several motivations for the present work: 
(i) a Lagrangian approach is more intuitive for particle-laden flows and therefore relevant; 
(ii) some specific issues are particularly arduous to tackle in the Eulerian two-fluid approach, requiring the Lagrangian one, namely, local but nonlinear phenomenon like polydispersity and chemical reactions \citep{Pope_turbulent,Fox2003}; 
(iii) the Lagrangian approach provides a more detailed information content. Notably, the velocity of the fluid seen by particles is available whereas it is not in two-fluid models;
(iv) the possibility to test different models will provide insights into the role of each term, notably the granular energy, in the development of the particle and fluid turbulence; 
(v) it is an intriguing perspective to consider the possible unification of the present Lagrangian stochastic model with the classical ones used to describe granular matter \citep{puglisi2014transport}.

To avoid errors and/or confusion, it is important to build the model through a coarse-graining approach starting from the fundamental description, as in classical statistical mechanics \citep{castiglione2008chaos}. We report here the main levels of description of our problem together with the main assumptions made when developing the Lagrangian model. 

(i) It is possible to neglect detailed molecular effects, since particles are always considered much larger than molecules (the diameter is $d_p \gg $ nm), so that the more fundamental level considered is hydrodynamic. 
Nonetheless,  when the particle diameter is $d_p \le \mu$m,  an effective Brownian term has to be added to take into account particle--solvent interactions. 
When their sizes are comparable, or larger, than the smallest active scale of the fluid flow (\eg the Kolmogorov dissipative scale in three-dimensional turbulence), determining particles dynamics requires fully resolving the fluid flow around them. 
This description is hence given by the Navier--Stokes (NS) equation for the fluid phase and by Newton's equations for particles, where the whole particle physics is represented through applying no-slip boundary conditions at the surface of each particle. 
Various numerical techniques have been developed to the end of studying finite-size particles, such as immersed boundaries  \citep{fadlun2000combined,lucci2010modulation}, two-fluid VOF \citep{moule2014highly} and level-set \citep{kwakkel2012efficient,moule2014highly} allowing one to reach volume fractions of the order of 2--40\%  \citep{ten2004fully,Chouippe2015,Picanoeal2015,Fornarietal2016,Tanaka2017}. 
This can be considered the \emph{microscopic} level of description of particle-laden flows.
However, the number of resolved particles is generally limited because of the high computational demand, and this level of description is often unnecessary since particles are small enough to justify a point-wise approximation \citep{Gat_83,Max_83}, even though finite-size effects are important for small systems \citep{pedley1992hydrodynamic}, and larger particles \citep{picano2013shear}.

(ii) Since the microscopic level is generally too detailed for realistic applications, it is tempting to search for a kinetic description of the particle phase, in analogy with the Boltzmann treatment of the molecules of a fluid \citep{cercignani1988boltzmann}. 
If the fluid presence can be neglected (a dense dry suspension), this is the standard problem of granular flows. Grains replace molecules as microscopic constituents and a kinetic equation can be written for a probability density function (pdf) $f(\mathbf{x},\mathbf{V}_p,t)$, where $\mathbf{x},\mathbf{V}_p$ represent the possible position and velocity of the grains. 
In principle, the approach is justified, yet the difficult issue here is to propose a suitable closure for the collision term, since grains are different from molecules and notably collisions are not necessarily elastic. 
In such a framework, the kinetic approach has been developed for rapid granular flows animated by elastic or inelastic collisions that drive the distribution function towards a local Maxwell--Boltzmann equilibrium \citep{Jen_83,Jen_85a,Lun_86,brey1998hydrodynamics,brilliantov2010kinetic}.

When the suspension is not dry, the fluid phase has to be added. 
If we consider that the fluid velocity at the position of each particle is known (from numerical simulations or analytical specification), the generalisation consists in specifying the force exerted by the fluid on particles, which is added as an external term in the kinetic equation, but the distribution function remains well defined as $f(\mathbf{x},\mathbf{V}_p,t)$.
The kinetic equation reads \citep{Jen_83}
\begin{equation}
\frac{\partial f}{\partial t} + \frac{\partial}{\partial {\bf x}} \bcdot ({\bf V}_p f) + \frac{\partial}{\partial {\bf V}_p} \bcdot [( \mathcal{A}_p + {\bf g})f] = \mathcal{C}
\label{eq:KT}
\end{equation}
where $\mathcal{A}_p$ is the acceleration due to fluid--particle interactions, ${\bf g}$ is the gravity acceleration and $\mathcal{C}$ is the collision operator. 
It is worth underlining that if the fluid field is not known, 
the problem is not well posed and the kinetic approach, that is only $\mathbf{x}$ and $\mathbf{V}_p$ are considered as variables, is incomplete \citep{minier2015kinetic}.
The kinetic level of description can be considered valid in a wide range of situations. In analogy with statistical mechanics terminology,
this is the mesoscopic level of description.

(iii) From the kinetic equation it is possible to derive corresponding hydrodynamic equations through averaging over the kinetic distribution function \citep{huang1963statistical}.
These equations are purely formal if a systematic procedure to compute averages is not given and the distribution function is unknown.
If one considers local equilibrium, notably the Maxwellian for elastic collisions, it is possible to resort to the Chapman--Enskog asymptotic method, valid for the kinetic theory of dilute gases \citep{chapman1970mathematical}.
First works derived hydrodynamic equations considering Maxwellian equilibrium and in absence of the fluid-phase force, yet small deviations from Maxwellian can be taken into account considering instead the Sonine polynomials \citep{van1998velocity,garzo2012enskog}.
Assuming that particles are frictionless hard spheres of equal density and diameter (i.e.\ monodisperse) and that collisions are nearly elastic, the conservation of mass and momentum of the hydrodynamic variables (zeroth and first-order moments of the kinetic distribution function), in the presence of a constant-density fluid, are given by the following equations:
\begin{equation}
\frac{\partial \alpha_p}{\partial t} + \bfnabla \bcdot \alpha_p {\bf U}_p = 0,
\label{eq:continuity_part}
\end{equation}
\begin{equation}
\frac{\partial \alpha_p {\bf U}_p}{\partial t} + \bfnabla \bcdot \alpha_p  ( {\bf U}_p \otimes {\bf U}_p +  {\bf P}) = \alpha_p \left(\frac{{\bf U}_f - {\bf U}_p}{\tau_p} + {\bf g} \right)
\label{eq:momentumKT}
\end{equation}
where  $\alpha_p$ is the particle-phase volume fraction, ${\bf U}_p$ is the particle-phase velocity,  ${\bf U}_f$ is the fluid-phase velocity and 
${\bf P}$ is the particle-phase pressure tensor, given by the second-order moments of the kinetic distribution function \citep{Jen_83}. 
${\tau_p}$ is the particle response-time, whose precise definition will be given shortly.
The gravity acceleration  ${\bf g}$ has been considered as external force.
The momentum exchange between the phases in \eqref{eq:momentumKT}, is due only to drag, since small, high-density particles are considered. From \eqref{eq:momentumKT}, the transport equation for the particle-phase velocity tensor product can be obtained as
\begin{equation}
\frac{\partial \alpha_p {\bf U}_p \otimes {\bf U}_p}{\partial t} + \bfnabla \bcdot (\alpha_p {\bf U}_p \otimes {\bf U}_p \otimes {\bf U}_p) + [{\bf U}_p \otimes \bfnabla \bcdot (\alpha_p {\bf P} )]^{\dagger} = \alpha_p \left[{\bf U}_p \otimes \left( \frac{{\bf U}_f - {\bf U}_p}{\tau_p} + {\bf g} \right) \right]^{\dagger}
\label{eq:tensorKT}
\end{equation}
where the symbol $[\bcdot]^\dagger$ implies the summation of a second-order tensor with its transpose.
For non-equilibrium flows a transport equation for the pressure tensor is necessary, and can be derived from \eqref{eq:KT} and \eqref{eq:tensorKT}:
\begin{equation}
\frac{\partial \alpha_p {\bf P}}{\partial t} + \bfnabla \bcdot \alpha_p ({\bf U}_p \otimes {\bf P} + {\bf Q}) = - \alpha_p ({\bf P} \bcdot \bfnabla {\bf U}_p )^{\dagger} - \frac{2}{\tau_p} \alpha_p {\bf P} + \frac{12}{\sqrt{\pi} d_p}\alpha_p^2 \Theta^{1/2} ( \Delta^* - {\bf P}).
\label{eq:pressure}
\end{equation}
In this equation $\Theta$ ($= \frac{1}{3}Tr({\bf P})$) is the granular temperature, ${\bf Q}$ is a heat-flux tensor that contains the third-order central moments of the velocity distribution function, and the last term on the right-hand side is the particle--particle collision term that has been closed using the Bhatnagar--Gros--Krook (BGK) approximation \citep{bhatnagar1954model} extended to inelastic collisions \citep{passalacqua2011quadrature}, where $0 \le e \le 1$ is the coefficient of restitution, $d_p$ is the particle diameter and $\Delta ^*$ is the second-order moments of the collisional equilibrium distribution, given by
\begin{equation}
\Delta ^* = \frac{1}{4}(1+e)^2 \Theta {\bf I} + \frac{1}{4}(1-e)^2 {\bf P}.
\label{eq:delta*}
\end{equation}
 By taking one-third of the trace of \eqref{eq:pressure}, the equation for the granular temperature can be found
\begin{equation}
\frac{\partial \alpha_p \Theta}{\partial t} + \bfnabla \bcdot \alpha_p \left( {\bf U}_p \Theta + \frac{2}{3} {\bf q} \right)= - \frac{2}{3} \alpha_p {\bf P} \colon \bfnabla {\bf U}_p - \frac{2}{\tau_p}\alpha_p \Theta - \frac{6 (1-e^2)}{\sqrt{\pi}d_p} \alpha_p^2 \Theta^{3/2}
\label{eq:granular_temp}
\end{equation}
where ${\bf q}$ is the granular temperature flux, i.e.\ the trace of ${\bf Q}$.
This is the \emph{hydrodynamic} level of description and is inherently macroscopic.

While a large spectra of conditions fall within this hydrodynamic framework, it has been nevertheless shown via numerical simulations and experiments that in several situations the local equilibrium approximation does not hold and the hydrodynamic equations display large errors \citep{goldhirsch1993clustering,du1995breakdown,kadanoff1999built,puglisi1999kinetic},
becoming rather formal.
In particular, when collisions are very inelastic and/or the flow is too dense, that is the volume fraction of the particle phase $\alpha_p \gtrsim 40\%$.
On the basis of this discussion, the present work can be surely considered valid for 
moderately dense rapid flows ($1\% \le \alpha_p \le 20\%$). Outside these limits, 
each case should be considered carefully, even though the hydrodynamic description may apply there as well.

(iv) We have seen that a statistical description arises from the deterministic microscopic one because of the effect of particle interactions.
Now it is also possible that the fluid-phase displays important fluctuations, that is turbulence, because of the sensitivity to initial conditions. If the flow is not too dense, this effect can be important and a new intriguing dynamics is triggered.
This means that the above description remains correct but only for a given single realisation of the fluid, that is for a single system or experiment. 
Different realizations will lead to a different fluid velocity and hence to a different instantaneous particle dynamics.
In this case, what is important are statistical observables.
 If one performs averages of the relevant observables over different realisations, the statistical quantities obtained are the Reynolds averaged (RA) ones. 
For this reason, the hydrodynamic level may be considered ``more microscopic" than the RA one.
This is why 
\cite{Fox2014} defined the hydrodynamic level as the mesocopic or mesoscale moment level to distinguish clearly it from the following statistically averaged. For simplicity, we prefer here to stick with the usual definition employed in statistical mechanics.

The exact RA transport equations may be derived taking the RA of the hydrodynamic equations \eqref{eq:continuity_part}--\eqref{eq:pressure}. 
To this aim, it is necessary to introduce phase-averaged (PA) quantities \citep{Fox2014}, which are defined as the Reynolds average (RA) weighted with the phase volume fraction, therefore for the fluid phase $\langle ( \bcdot)\rangle_f = \langle \alpha_f (\bcdot )\rangle / \langle \alpha_f \rangle$ and for the particle phase $\langle ( \bcdot)\rangle_p = \langle \alpha_p (\bcdot )\rangle / \langle \alpha_p \rangle$. The decomposition adopted for the fluid- and particle-phase velocities with respect to the PA operator is the following:
\begin{equation}
{\bf U}_f = \langle {\bf U}_f \rangle_f + {\bf u}_f,
\label{eq:decomp}
\end{equation}
\begin{equation}
{\bf U}_p = \langle {\bf U}_p \rangle_p + {\bf u}_p
\label{eq:decomp2}
\end{equation}
with $\langle {\bf u}_f \rangle_f = 0$ and $\langle {\bf u}_p \rangle_p = 0$, but $\langle {\bf u}_f \rangle_p \neq 0$.
Taking the RA of \eqref{eq:continuity_part} yields 
\begin{equation}
\frac{\partial \langle \alpha_p \rangle }{\partial t} + \bfnabla \bcdot \langle \alpha_p \rangle \langle {\bf U}_p \rangle_p = 0 .
\label{eq:particle_cont_RA}
\end{equation}
The PA particle-phase momentum equation found from \eqref{eq:momentumKT} is given by
\begin{align}
\frac{\partial \langle \alpha_p \rangle \langle {\bf U}_p \rangle_p}{\partial t} & + \bfnabla \bcdot \langle \alpha_p \rangle ( \langle {\bf U}_p \rangle_p \otimes \langle {\bf U}_p \rangle_p  + \langle \mathcal{P} \rangle _p) =   \langle \alpha_p \rangle  \left( \left\langle \frac{ {\bf U}_f - {\bf U}_p }{\tau_p} \right\rangle_p +  {\bf g} \right)
\label{eq:particle_mom_RA}
\end{align} 
where $\langle \mathcal{P} \rangle_p = \langle {\bf P} \rangle_p + \langle {\bf u}_p \otimes {\bf u}_p \rangle_p$ is the sum of the particle-phase stress tensor and the particle-phase Reynolds stress tensor.

The PA particle-phase stress tensor that appears in  $\langle \mathcal{P} \rangle_p$ is governed by the following equation found from \eqref{eq:pressure}:
\begin{multline}
\frac{\partial \langle \alpha_p \rangle \langle {\bf P} \rangle_p}{\partial t} + \bfnabla \bcdot \langle \alpha_p \rangle ( \langle {\bf U}_p \rangle_p \otimes \langle {\bf P} \rangle_p + \langle {\bf u}_p \otimes {\bf P} \rangle_p + \langle {\bf Q}\rangle_p ) = \\  
- \langle \alpha_p  \rangle (\langle {\bf P} \rangle_p \bcdot \bfnabla \langle {\bf U}_p \rangle_p + \langle {\bf P} \bcdot \bfnabla {\bf u}_p \rangle_p )^{\dagger} - \frac{2}{\tau_p} \langle \alpha_p \rangle \langle {\bf P} \rangle_p + \frac{12 \langle \alpha_p \rangle}{\sqrt{\pi} d_p} \langle \alpha_p \Theta^{1/2} ( \Delta^* - {\bf P}) \rangle_p.
\label{eq:pressure_RA}
\end{multline}
Taking the RA of the granular temperature transport equation \eqref{eq:granular_temp} (or one-third the trace of \eqref{eq:pressure_RA}) yields
\begin{multline}
\frac{\partial \langle \alpha_p \rangle \langle \Theta \rangle_p}{\partial t} + \bfnabla \bcdot \langle \alpha_p \rangle \left( \langle {\bf U}_p \rangle_p  \langle \Theta \rangle_p + \langle {\bf u}_p \Theta \rangle_p + \frac{2}{3} \langle {\bf q}\rangle_p \right) \\ 
= - \frac{2}{3} \langle \alpha_p  \rangle (\langle {\bf P} \rangle_p \colon \nabla \langle {\bf U}_p \rangle_p + \langle {\bf P} \colon \bfnabla {\bf u}_p \rangle_p ) - \frac{2}{\tau_p} \langle \alpha_p \rangle \langle \Theta \rangle_p - \frac{6 (1 - e^2) }{\sqrt{\pi} d_p} \langle \alpha_p^2 \Theta^{3/2} \rangle.
\label{eq:granular_RA}
\end{multline}
The transport equation for the particle-phase Reynolds stress tensor is computed by subtracting the transport equation for the particle-phase mean velocity tensor product from the RA of \eqref{eq:tensorKT}, yielding
\begin{multline}
\frac{\partial \langle \alpha_p \rangle \langle {\bf u}_p \otimes {\bf u}_p \rangle_p}{\partial t}  + \bfnabla \bcdot \langle \alpha_p \rangle ( \langle {\bf U}_p \rangle_p \otimes \langle {\bf u}_p \otimes {\bf u}_p \rangle_p + \langle {\bf u}_p \otimes {\bf u}_p \otimes {\bf u}_p \rangle_p + \langle {\bf P} \otimes {\bf u}_p \rangle_p^{\dagger}) = \\ 
- \langle \alpha_p \rangle (\langle {\bf u}_p \otimes {\bf u}_p \rangle_p \bcdot \bfnabla \langle {\bf U}_p \rangle _p )^{\dagger} + \langle \alpha_p \rangle \langle {\bf P} \bcdot \bfnabla {\bf u}_p \rangle_p^{\dagger} +  \frac{\langle \alpha_p \rangle}{\tau_p}( \langle {\bf u}_f \otimes {\bf u}_p \rangle_p - \langle {\bf u}_p \otimes {\bf u}_p \rangle_p )^{\dagger}. 
\label{eq:particle_reynolds_RA}
\end{multline}
It is worth noting that the turbulent dissipation in \eqref{eq:particle_reynolds_RA}, $\langle {\bf P} \bcdot \bfnabla {\bf u}_p \rangle_p$, appears as a source term for $\langle {\bf P} \rangle$ in \eqref{eq:pressure_RA}, therefore it can be expressed in both equations as a particle-phase dissipation tensor
\begin{equation}
\bds{\varepsilon}_{p} = \langle {\bf P} \bcdot \bfnabla {\bf u}_p \rangle_p.
\label{eq:dissip-tensor-prop}
\end{equation} 
This term, and several others, like triple correlations and fluxes, are evidently unclosed, which prevents us from solving the above set of RA equations. 
Recently, a model for the closure of these equations has been proposed \citep{capecelatro2016}. 

\begin{figure}
\begin{center}
{\includegraphics[width=.95\textwidth]{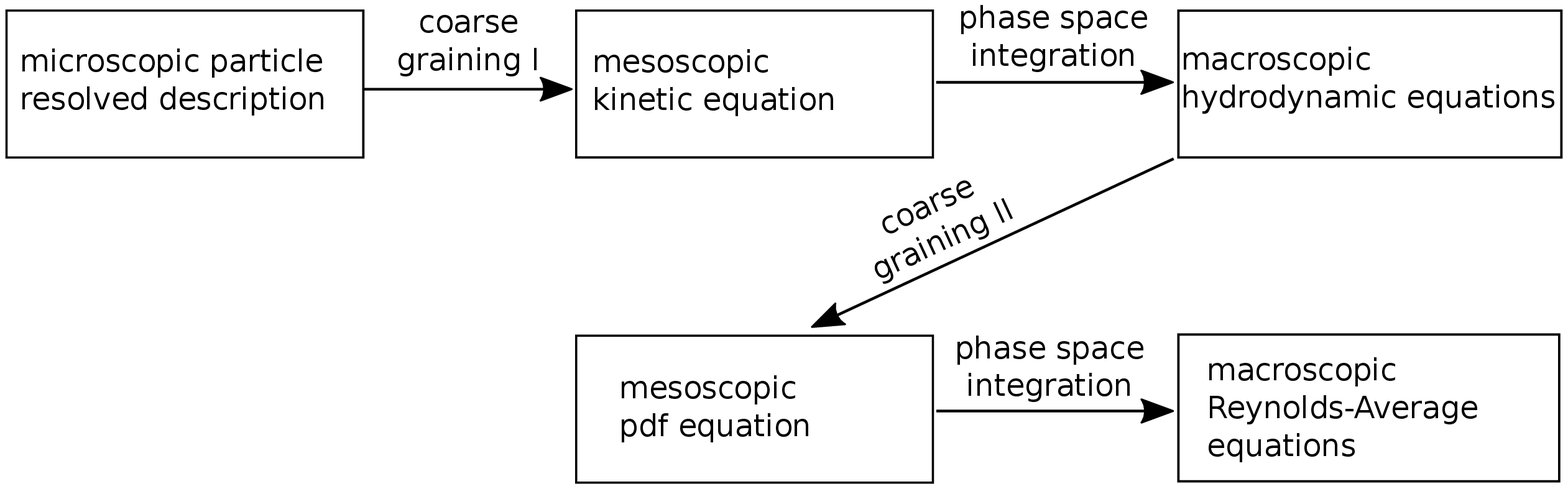}}
\caption{{Levels of description of the fluid--particle system. The focus of this work is to develop a Lagrangian pdf model that is equivalent to the mesoscopic pdf equation.}}
\label{coarsening}
\end{center}
\end{figure}

The resulting picture obtained through the coarse graining is sketched in figure~\ref{coarsening}.
In the following, we shall develop the equivalent coarse-graining procedure using a Lagrangian approach. 
In the end, a hybrid Eulerian/Lagrangian method will be available to simulate moderately dense particle-laden flows, in which the role of collisions will be taken into account statistically. 
An explicit modelling of the correlated and uncorrelated components of particle velocity will be carried out.
In particular, we will propose a closed Lagrangian stochastic model for moderately dense particle-laden flows  that gives RA equations very similar to the exact ones presented above, first derived in \cite{Fox2014}.
\section{The microscopic equations}
\label{sec:meso}
\subsection{Fluid phase}
Hydrodynamic equations for the fluid phase are obtained by applying a volume-filtering operator to the microscale NS equations  \citep{anderson1967fluid}, which for a constant-density fluid gives
\begin{equation}
\frac{\partial \alpha_f }{\partial t} + \bfnabla \bcdot \alpha_f {\bf U}_f = 0
\label{eq:fluid_cont}
\end{equation}
and
\begin{equation}
\frac{\partial \alpha_f {\bf U}_f}{\partial t} + \bfnabla \bcdot \alpha_f {\bf U}_f \otimes {\bf U}_f = - \frac{1}{\rho_f} \bfnabla p_f + \frac{1}{\rho_f} \bfnabla \bcdot \bds{\sigma}_f - \alpha_f \Phi \frac{{\bf U}_f - {\bf U}_p}{\tau_p} + \alpha_f {\bf g}
\label{eq:fluid_NS}
\end{equation} 
where ${\bf U}_f$ and $p_f$ are the instantaneous fluid-phase velocity and pressure, $\alpha_f$ is the fluid-phase volume fraction, $\rho_f$ and $\nu_f$ are the fluid-phase density and kinematic viscosity and ${\bf g}$ the acceleration due to gravity. 
The particle characteristic time scale $\tau_p$ is defined as
\begin{equation}
\tau_p = \frac{\rho_p d_p^2}{18 \rho_f \nu_f}
\label{eq:def_taup}
\end{equation}
with $\rho_p$ and $d_p$ being, respectively, the particle-phase density and diameter. 
The (instantaneous) mass loading $\Phi$ is defined as
\begin{equation}
\Phi = \frac{\rho_p \alpha_p}{\rho_f \alpha_f} .
\end{equation}
The fluid-phase viscous stress tensor is defined as
\begin{equation}
\bds{\sigma}_f = \rho_f \nu_f [\bfnabla {\bf U}_f + (\bfnabla {\bf U}_f)^T - \frac{2}{3} \bfnabla \bcdot {\bf U}_f {\bf I}]
\end{equation}
where ${\bf I}$ is the identity tensor.
The unclosed terms coming from the volume filtering of the microscopic stress tensor have been neglected here since it has been shown that they do not influence noticeably the flow physics \citep{Fox2015}. 
Moreover, concerning the effects of the particles on the fluid, we have retained only the drag force, since we are considering cases where $\rho_p/\rho_f \gg 1$.

From \eqref{eq:fluid_NS} and enforcing \eqref{eq:fluid_cont}, a transport equation for the fluid-phase velocity tensor product can be found
\begin{multline}
\frac{\partial \alpha_f  {\bf U}_f \otimes {\bf U}_f}{\partial t}  + \bfnabla \bcdot (\alpha_f {\bf U}_f \otimes {\bf U}_f \otimes {\bf U}_f) = 
- \frac{1}{\rho_f}({\bf U}_f \otimes \bfnabla p_f)^{\dagger} + \frac{1}{\rho_f} ({\bf U}_f \otimes \bfnabla \bcdot \bds{ \sigma}_f)^{\dagger} \\
- \alpha_f \Phi \left[ {\bf U}_f \otimes \left( \frac{{\bf U}_f - {\bf U}_p}{\tau_p} \right) \right]^{\dagger} + \alpha_f ({\bf U}_f \otimes {\bf g})^{\dagger}.
\end{multline}

\subsection{Particle phase}
\label{sec:particle_phase}
As anticipated in the introduction, we neglect the finite-size effect and hence the displacement of a point-wise particle is described by the following Newton's second law of motion \citep{Gat_83,Max_83}
\begin{equation}
\frac{d {\bf V}_p^{(k)}}{dt}= \frac{{\bf U}_f[{\bf x}_p^{(k)}] - {\bf V}_p^{(k)}}{\tau_p} + {\bf F}_c^{(k)} + {\bf g} \label{eq:maxey_riley}
\end{equation}
where ${\bf x}_p^{(k)}$ is the centre position of particle $k$ and ${\bf F}_c$ is the collisional acceleration experienced by the particle. 
Moreover, since it is assumed that $\rho_p \gg \rho_f$, only the drag force exerted by the fluid is considered, while all other contributions from the fluid phase (\eg added mass and lift forces) are neglected.

\section{Mesoscopic equations} \label{sec:collisions}

As anticipated in the introduction, it is in general possible to 
go through a statistical kinetic description for particle position and velocity,
which reads
\begin{equation}
\frac{\partial f}{\partial t} + \frac{\partial}{\partial {\bf x}} \bcdot ({\bf V}_p f) + \frac{\partial}{\partial {\bf V}_p} \bcdot [( \mathcal{A}_p + {\bf g})f] = \mathcal{C}~.
\end{equation}
This is the kinetic mesoscopic level in figure~\ref{coarsening}.
It is important to underline that this picture is meaningful whenever the fluid realisation is given. For laminar flows, which are not sensitive to initial and boundary conditions, the flow is always the same, independent from the realisation, given the geometry and the initial state. 
For turbulent flows, the situation is more complicated and, because of the inevitable presence of small perturbations, the instantaneous field changes at each realisation. 
For this reason, \cite{fevrier2005partitioning} correctly point out that $f$ is a probability density function conditioned on the fluid-flow realisation and should be noted as $f(\mathbf{x},\mathbf{V}_p,t)=\tilde{f}(\mathbf{x},\mathbf{V}_p,t \vert \mathcal{H}_f)$, where $\mathcal{H}_f$ indicates a given fluid-flow realisation. This point being clarified, we stick to the standard notation for the sake of simplicity and clarity.

Since we are considering particle flows in the collisional regime,  inter-particle collisions should be taken into account. In particular, \cite{Fox2015} and \cite{capecelatro2016} pointed out that it is of crucial importance for the modelling, the separation of the particle velocity into a spatially correlated component and an uncorrelated one, as initially introduced by \cite{fevrier2005partitioning} in the dilute case. 
For example, numerical simulations indicate that in cluster-induced turbulence (CIT) the granular temperature is no longer negligible with respect to the turbulent kinetic energy \citep{Fox2015}. 

We have seen that in the Eulerian framework the hydrodynamic equations can be obtained by integrating over $f(\mathbf{x},\mathbf{V}_p,t)$,  \eqref{eq:continuity_part}--\eqref{eq:granular_temp}.  Notably 
the macroscopic velocity is defined as
$\mathbf {U}_p[{\bf x},t]=\int \mathbf{V}_p f(\mathbf{x},\mathbf{V}_p,t) d\mathbf{V}_p$, always conditioned on a given fluid-flow realization. 
To obtain its Lagrangian evolution equation, we can introduce a Lagrangian coarse-graining operator $f^L({\bf x}^{(k)}(t),{\bf V}_p,t)$ to be applied to the Lagrangian transport equation of the particle velocity ${\bf V}_p^{(k)}$. This operator is a local average and yields a decomposition that would lead to the same moment equations of \eqref{eq:continuity_part}--\eqref{eq:pressure}, since it is simply the Lagrangian projection of the kinetic distribution function.
The operator has the following properties:
(i) the residual part has zero mean;
(ii) the residual part is uncorrelated with the filtered part.
With this definition, we have that the hydrodynamic or correlated component is
\begin{equation}
\alpha_p {\bf U}_p[{\bf x} = {\bf x}^{(k)}(t),t] 
= 
 \widetilde{\bf V}_{p} ({\bf x}^{(k)}(t),t) =\int {\bf V}_p f^L({\bf x}^{(k)}(t),{\bf V}_p,t) \, d{\bf V}_p ,
\end{equation}
and the complete velocity is given by
\begin{equation}
{\bf V}_p^{(k)} =  { \bf \widetilde{V}}_{p} ({\bf x}^{(k)}(t),t) + \bds{\delta} {\bf V}_p^{(k)} = 
\alpha_p {\bf U}_p[{\bf x} = {\bf x}^{(k)}(t),t] + \bds{\delta} {\bf V}_p^{(k)}
\end{equation}
where we have indicated the coarse-grained quantities also with the tilde symbol \, $\widetilde{}$ \,, because it simplifies the notation for the manipulation of the equations which follow.
It is interesting to remark that this coarse graining shares some similarity with a recent Lagrangian large-eddy simulation filtering formalism applied to particle-laden flows \citep{innocenti2016lagrangian}.

As usual in Lagrangian approaches \citep{Pop_85,Min_01}, the purpose is to propose Lagrangian equations that transport the pdf.
To get such equations, we apply the coarse graining to the Lagrangian equation of the particle motion in which we consider only drag and gravity, neglecting for the moment the explicit effect of collisions:
\begin{equation}
\widetilde{\frac{d V_{p,i}}{dt}} =
\frac{\partial \widetilde{V}_{p,i}}{\partial t} + \widetilde{V}_{p,j} \frac{\partial \widetilde{V}_{p,i}}{\partial x_j} + \frac{\partial \widetilde{\delta V_{p,i} \delta V_{p,j}}}{\partial x_j} 
= \frac{\widetilde{U}_{f,i} - \widetilde{V}_{p,i}}{\tau_p} + g_i ,
\label{eq:filt1}
\end{equation}
which is the Lagrangian equivalent of \eqref{eq:momentumKT}. 
The Lagrangian transport equation of $\widetilde{V}_{p,i}$ can therefore be expressed as
\begin{align}
\frac{d \widetilde{V}_{p,i}}{dt} & = \frac{\partial \widetilde{V}_{p,i}}{\partial t} + V_{p,j} \frac{\partial \widetilde{V}_{p,i}}{\partial x_j} \notag \\
& = \frac{\partial \widetilde{V}_{p,i}}{\partial t} + \widetilde{V}_{p,j} \frac{\partial \widetilde{V}_{p,i}}{\partial x_j} + \delta V_{p,j} \frac{\partial \widetilde{V}_{p,i}}{\partial x_j} \notag \\ & = \frac{\widetilde{U}_{f,i} - \widetilde{V}_{p,i}}{\tau_p} + g_i - \frac{\partial \widetilde{\delta V_{p,i} \delta V_{p,j}}}{\partial x_j} + \delta V_{p,j} \frac{\partial \widetilde{V}_{p,i}}{\partial x_j}.
\label{eq:filt_lag1}
\end{align}

On the other hand, it is also possible to explicitly express the residual part,
since its material derivative is given by
\begin{align}
\frac{d \delta V_{p,i}}{dt} & = \frac{d ( V_{p,i} - \widetilde{V}_{p,i} )}{dt} \notag \\ &= \frac{\widetilde{U}_{f,i} - V_{p,i}}{\tau_p} + g_i - \frac{\widetilde{U}_{f,i} - \widetilde{V}_{p,i}}{\tau_p} - g_i + \frac{\partial \widetilde{\delta V_{p,i} \delta V_{p,j}}}{\partial x_j} - \delta V_{p,j} \frac{\partial \widetilde{V}_{p,i}}{\partial x_j} \notag \\  & = - \frac{\delta V_{p,i}}{\tau_p} + \frac{\partial \widetilde{\delta V_{p,i} \delta V_{p,j}}}{\partial x_j} - \delta V_{p,j} \frac{\partial \widetilde{V}_{p,i}}{\partial x_j} .
\end{align}
The Lagrangian transport equation for the uncorrelated energy tensor, i.e.\ the particle-phase pressure tensor, will be
\begin{multline}
d (\delta V_{p,i} \delta V_{p,j} ) = -2 \frac{\delta V_{p,i} \delta V_{p,j}}{\tau_p} dt - \delta V_{p,i} \delta V_{p,k} \frac{\partial \widetilde{V}_{p,j}}{\partial x_k} dt - \delta V_{p,j} \delta V_{p,k} \frac{\partial \widetilde{V}_{p,i}}{\partial x_k} dt \\ 
+ \delta V_{p,i} \frac{\partial \widetilde{\delta V_{p,j} \delta V_{p,k}}}{\partial x_k} dt + \delta V_{p,j} \frac{\partial \widetilde{\delta V_{p,i} \delta V_{p,k}}}{\partial x_k} dt .
\label{eq:deltavdeltav}
\end{multline}
Following this decomposition, we can thus define two different Lagrangian processes, one for the coarse-grained particle velocity and one for the residual component:
\begin{eqnarray}
\left\{
\begin{split}
& d \widetilde{V}_{p,i} = \frac{\widetilde{U}_{f,i} - \widetilde{V}_{p,i}}{\tau_p} dt + g_i dt - \frac{\partial \widetilde{\delta V_{p,i} \delta V_{p,k}}}{\partial x_k} dt + \delta V_{p,k} \frac{\partial \widetilde{V}_{p,i}}{\partial x_k} dt \\
& d \delta V_{p,i} = - \frac{\delta V_{p,i}}{\tau_p} dt + \frac{\partial \widetilde{\delta V_{p,i} \delta V_{p,k}}}{\partial x_k} dt - \delta V_{p,k} \frac{\partial \widetilde{V}_{p,i}}{\partial x_k} dt
\end{split}
\right.
\label{eq:exact_Lag}
\end{eqnarray}
These equations constitute the \emph{Lagrangian mesoscopic equations}, and contain the  hydrodynamic or \emph{macroscopic} equation \eqref{eq:continuity_part}--\eqref{eq:pressure}, except for the collision part that will be treated shortly.
The  \emph{macroscopic} equations can be obtained by applying the coarse graining on the relevant observables.
Notably, the particle-phase pressure tensor is given by $P_{ij} = \widetilde{\delta V_{p,i} \delta V_{p,j}}$.

\section{Macroscopic Reynolds-average equations of motion}
\label{sec:macro}
Keeping in mind the definitions about the phase-average (PA) given in \S \ref{sec:introduction}, see \eqref{eq:decomp} and \eqref{eq:decomp2}, we recall that the following identity holds between PA and RA of a Lagrangian quantity:
\begin{equation}
 \langle {\bf U}_p \rangle _p  = \langle \widetilde{{\bf V}}_p \rangle = \langle {\bf V}_p \rangle ,
\label{eq:rel-PA_Lag}
\end{equation}
since, while doing a Lagrangian average, we shift from a Lagrangian to an Eulerian description by means of local averages, and thus implicitly weighting the phase with its volume fraction. The last equality in \eqref{eq:rel-PA_Lag} comes from property (i) of the Lagrangian coarse-graining operator $\langle \delta {\bf v}_p \rangle = 0$.

Adopting, from now on, the Lagrangian formalism, the total particle-phase fluctuating energy is defined by
\begin{equation}
\kappa_p = \frac{1}{2} \langle {\bf v}_p \bcdot {\bf v}_p \rangle
\end{equation}
where ${\bf v}_p = {\bf V}_p - \langle {\bf V}_p \rangle$ is the total fluctuation in the particle velocity, which can also be expressed as the sum of two contributions: the fluctuation of the coarse-grained, correlated part, $\widetilde{{\bf v}}_p = \widetilde{{\bf V}}_p - \langle \widetilde{{\bf V}}_p \rangle$, and the uncorrelated part, $\delta {\bf V}_p$.
By means of this decomposition, the total particle-phase fluctuating energy can, in turn, be split in two contributions, the turbulent particle-phase kinetic energy and the granular temperature:
\begin{equation}
\kappa_p = k_p + \frac{3}{2} \langle \Theta_p \rangle
\end{equation}
where
\begin{equation}
k_p = \frac{1}{2} \langle \widetilde{{\bf v}}_p \bcdot \widetilde{{\bf v}}_p \rangle \qquad \qquad \langle \Theta \rangle_p = \frac{1}{3} \langle \widetilde{ \delta{\bf V}_p \bcdot  \delta{\bf V}_p } \rangle .
\end{equation}
It is worth remarking that the turbulent particle-phase kinetic energy can also be expressed via the PA as $k_p =  \frac{1}{2} \langle {\bf u}_p \bcdot {\bf u}_p \rangle_p$. Concerning the granular temperature, it can equally be found from the trace of the particle-phase pressure tensor 
\begin{equation}
\langle {\bf P} \rangle = \langle \widetilde{\delta {\bf V}_p \otimes \delta{\bf V}_p} \rangle.
\end{equation}
The distinction between $k_p$ and $\langle \Theta_p \rangle$ is crucial in turbulence modeling of multiphase flows because, for example, they have different boundary conditions and the particle--particle collision frequency depends on $\Theta_p$  \citep{capecelatro2016}.

\subsection{Exact equations}
When the flow is turbulent, changing slightly the realization of the fluid phase provokes a large difference also in the particle dynamics.
This means that only observables averaged over many fluid realizations (or over time if the system is statistically stationary and assumed ergodic) 
are relevant.
The wide range of length and time scales associated with turbulent multiphase flows makes a direct solution of the transport equations presented thus far intractable for most applications. Therefore it is necessary to 
solve directly the equations for statistical observables.

From the equation for $\widetilde{{\bf V}}_{p}$  \eqref{eq:exact_Lag}, applying a RA (ensemble averaging over a large number of fluid realizations), the exact RA equations can be retrieved for the particle mean velocity $\langle {\bf V}_p \rangle = \langle \widetilde{{\bf V}}_p \rangle$ 
\begin{equation}
\frac{\partial \langle \alpha_p \rangle \langle \widetilde{{\bf V}}_p \rangle}{\partial t}  + \bfnabla \bcdot  \langle \alpha_p \rangle ( \langle \widetilde{{\bf V}}_p \rangle \otimes \langle \widetilde{{\bf V}}_p \rangle  + \langle \mathcal{P} \rangle) =  \langle \alpha_p \rangle   \left( \left\langle \frac{ \widetilde{{\bf U}}_f - \widetilde{{\bf V}}_p }{\tau_p} \right\rangle +  {\bf g} \right)~.
\end{equation} 
The Reynolds stresses of the correlated part $\langle \widetilde{{\bf v}}_p \otimes \widetilde{{\bf v}}_p \rangle$ can be obtained analogously,
and it gives the following Eulerian transport equation: 
\begin{eqnarray}
\frac{\partial \langle \alpha_p \rangle \langle \widetilde{{\bf v}}_{p} \otimes \widetilde{{\bf v}}_{p} \rangle}{\partial t} &+&\bfnabla \bcdot \langle \alpha_p \rangle (\langle \widetilde{{\bf V}}_{p} \rangle \otimes \langle \widetilde{{\bf v}}_{p} \otimes \widetilde{{\bf v}}_{p} \rangle) + \bfnabla \bcdot \langle \alpha_p \rangle \langle \widetilde{{\bf v}}_{p} \otimes \widetilde{{\bf v}}_{p} \otimes \widetilde{{\bf v}}_{p} \rangle = \notag \\ & - & \langle \alpha_p \rangle (\langle \widetilde{{\bf v}}_{p} \otimes \widetilde{{\bf v}}_{p} \rangle \bcdot \bfnabla  \langle \widetilde{{\bf V}}_{p} \rangle)^{\dagger}  - \langle \alpha_p \rangle  \langle \widetilde{{\bf v}}_{p} \otimes  \bfnabla \bcdot (\widetilde{\delta {\bf V}_{p} \otimes \delta {\bf V}_{p}}) \rangle^{\dagger}  \notag \\ & & \qquad \qquad \qquad \qquad \quad + \langle \alpha_p \rangle \frac{( \langle \widetilde{{\bf u}}_{f} \otimes \widetilde{{\bf v}}_{p} \rangle  -  \langle \widetilde{{\bf v}}_{p} \otimes \widetilde{{\bf v}}_{p} \rangle )^{\dagger}}{\tau_p} .
\label{eq:upup_filter2}
\end{eqnarray} 
Finally, applying first the coarse-graining operator and then the RA one to  \eqref{eq:deltavdeltav}, we get the equation for the particle-phase pressure tensor:
\begin{eqnarray}
\frac{\partial \langle \alpha_p \rangle \langle {\bf P} \rangle}{\partial t} &+& \bfnabla \bcdot \langle \alpha_p \rangle \left( \langle \widetilde{{\bf V}}_{p} \rangle \otimes \langle {\bf P} \rangle + \langle \widetilde{\delta {\bf V}_{p} \otimes \delta {\bf V}_{p}} \otimes \delta {\bf V}_{p} \rangle  + \langle \widetilde{\delta {\bf V}_{p} \otimes \delta {\bf V}_{p}} \otimes \widetilde{{\bf v}}_{p} \rangle \right)= \notag \\ 
&  & \qquad \qquad  - 2 \langle \alpha_p \rangle \frac{\langle {\bf P} \rangle}{\tau_p} - \langle \alpha_p \rangle (\langle {\bf P} \rangle \bcdot \bfnabla \langle \widetilde{{\bf V}}_p \rangle)^{\dagger}  -  \langle \alpha_p \rangle \langle {\bf P} \bcdot \bfnabla \widetilde{{\bf v}}_p \rangle^{\dagger} ~.
\label{eq:P_filter2}
\end{eqnarray} 
These RA macroscopic equations are exact but unclosed, and they are to be compared to \eqref{eq:particle_mom_RA}--\eqref{eq:particle_reynolds_RA}. 
This derivation demonstrates that both routes give the same equations and are therefore equivalent when collisions are neglected.

\subsection{Modelled equations}
To obtain a closed form of the macroscopic RA equations in Lagrangian terms, we
model directly the mesoscopic equations \eqref{eq:exact_Lag}. 
The modelling consists in replacing the ``faster'' terms with a stochastic model, and to close slow unclosed terms. From the modelled equations, RA quantities can be found as statistical moments of the underlying pdf. The terms to be replaced are those that imply a spatial gradient. In particular, the mean parts of those terms are retained, while fluctuations are modelled. 
In the following we have used the notation introduced at the beginning of the paper, therefore ${\bf U}_p$ stands for the model of $\widetilde{{\bf V}}_p$, while $\delta{\bf v}_p$ is the model for the residual component of the notional particles and must not be confused with the microscale fluctuations of a single particle. It is worth recalling that ${\bf U}_p$ and $\delta{\bf v}_p$ are two stochastic processes modelling two parts of the same quantity, therefore they are both advected by the particle total velocity, ${\bf V}_p = {\bf U}_p + \delta{\bf v}_p$.

The proposal for a stochastic model (still neglecting collisions) reads
\begin{align}
d \, { U}_{p,i}  =& \frac{{ U}_{s,i} - { U}_{p,i}}{\tau_p} \, dt + { g}_i \, dt - \frac{1}{\langle \alpha_p \rangle \rho_p} \frac{\partial \langle \alpha_p \rangle \rho_p \langle P_{ij} \rangle}{\partial x_j} + {\delta} { v}_{p,j} \frac{\partial \langle { U}_{p,i} \rangle}{\partial x_j} \, dt \notag \\ 
& - \frac{1}{T_{Lp}} ({ U}_{p,i} - \langle { U}_{p,i} \rangle ) \, dt + \sqrt{ C_{p}  \varepsilon_p} \; d{ W}_{p,i}  , \label{eq:up-final}\\
d \, {\delta} { v}_{p,i}  = & - \frac{{\delta} { v}_{p,i}}{\tau_p} \, dt  + \frac{1}{\langle \alpha_p \rangle \rho_p} \frac{\partial \langle \alpha_p \rangle \rho_p \langle P_{ij} \rangle}{\partial x_j} - {\delta} { v}_{p,j} \frac{\partial \langle { U}_{p,i} \rangle}{\partial x_j} \, dt 
 + B_{\delta,ij} \; d{ W}_{\delta,j}  \label{eq:dvp-final1}
\end{align}
where $k_p =   \frac{1}{2} \langle {\bf u}_{p} \bds{\bcdot} {\bf u}_{p} \rangle$ is the particle-phase turbulent kinetic energy, $\langle P_{ij} \rangle = \langle \delta v_{p,i} \delta v_{p,j} \rangle$ is the particle-phase pressure tensor and $\varepsilon_p$ represents the particle-phase dissipation, and has still to be specified. ${\bf U}_s$ is the fluid velocity seen by the particle, i.e.\ at the particle position, and its Lagrangian model will be defined in the following. The particle Lagrangian timescale is defined by
\begin{equation}
T_{Lp} = \left( \frac{1}{2} + \frac{3}{4} C_{0p} + \frac{f_{s}}{2} \right)^{-1} \frac{k_p}{\varepsilon_p}~,
\end{equation}
and the constant $C_p$ in the diffusion coefficient (\ref{eq:up-final}) is related to the Lagrangian timescale  to obtain a redistribution tensor and a dissipation tensor, by the relation
\begin{equation}
C_p = C_{0p} + \frac{2}{3} f_s .
\end{equation}
The parameter $f_s$ is introduced to account for anisotropy in the particle-phase dissipation tensor \citep{capecelatro2016}, which is needed to predict the anisotropy of the particle-phase pressure tensor $\langle {\bf P} \rangle$.

When the correlation $\langle \delta v_{p,i}  \delta v_{p,j} \rangle$ is evaluated, the diffusion matrix $B_{\delta}$ must give the particle-phase Reynolds-stress multiplied by the proper coefficient added to a diagonal isotropic part. Using the Choleski decomposition (see Appendix \ref{app:choleski}) we obtain:
\begin{align}
& B_{\delta,11} = \left(f_s \frac{\varepsilon_p}{k_p} \langle u_{p,1} u_{p,1} \rangle + (1-f_s) \frac{2}{3} \varepsilon_p \right)^{1/2},  \notag \\
& B_{\delta,i1} = \frac{1}{B_{\delta,11}} \left( f_s \frac{\varepsilon_p}{k_p} \langle u_{p,i} u_{p,1} \rangle \right), \quad  1 < i \le 3 \notag \\
& B_{\delta,ii} = \left( f_s \frac{\varepsilon_p}{k_p} \langle u_{p,i} u_{p,i} \rangle + (1-f_s) \frac{2}{3} \varepsilon_p - \sum_{j=1}^{i-1} B_{\delta,ij}^2 \right)^{1/2}, \quad 1 < i \le 3 \notag \\
& B_{\delta,ij} = \frac{1}{B_{\delta,jj}} \left( f_s \frac{\varepsilon_p}{k_p} \langle u_{p,i} u_{p,j} \rangle - \sum_{k=1}^{j-1} B_{\delta,ik} B_{\delta,jk} \right), \quad 1 < j < i \le 3 \notag \\
& B_{\delta,ij} = 0, \quad i < j \le 3 \, 
\end{align}
where repeated indices do not imply summation.

To complete the particle model we need to add explicitly the effect of collisions, i.e.\ the last term in \eqref{eq:pressure_RA}, which is due to the collisional equilibrium distribution function. In an Eulerian sense we can choose to model it as done by \cite{capecelatro2016}:
\begin{equation}
\langle \alpha_p \Theta_p^{1/2} ( \Delta^* - {\bf P}) \rangle = C_c \langle \alpha_p \rangle \langle \Theta_p \rangle^{1/2} (\langle \Delta^* \rangle - \langle {\bf P} \rangle )
\label{eq:model_coll_eul}
\end{equation}
where $\Delta^*$ is defined by \eqref{eq:delta*}, with which the last term of \eqref{eq:model_coll_eul} can be rewritten as
\begin{equation}
\langle \Delta^* \rangle - \langle {\bf P} \rangle = \frac{1}{4} (1+ e)^2 \langle \Theta_p \rangle {\bf I} - \frac{1}{4} (1+e) (3-e) \langle {\bf P} \rangle.
\end{equation}
Defining the characteristic time for collisions by $\tau_c = \sqrt{\pi} d_p / ( 6 C_c \langle \alpha_p \rangle \langle \Theta_p \rangle^{1/2} )$, we can express the collision tensor in the following form:
\begin{equation}
\mathcal{C}_{ij} = \frac{1}{2 \tau_c} [(1+e)^2 \langle \Theta_p \rangle \delta_{ij} - (1+e)(3-e)\langle P_{ij} \rangle ].
\end{equation} 
For elastic collisions $(e=1)$, $\mathcal{C}$ has zero trace. In order to obtain this collision tensor in the Eulerian RA equation of the particle-phase pressure tensor, two new terms have to be added in the stochastic equation for $\delta{\bf v}_p$  \eqref{eq:dvp-final1}, and the resulting model is
\begin{align}
d \, {\delta} { v}_{p,i} & = - \frac{{\delta} { v}_{p,i}}{\tau_p} \, dt + \frac{1}{\langle \alpha_p \rangle \rho_p} \frac{\partial \langle \alpha_p \rangle \rho_p \langle P_{ij} \rangle}{\partial x_j} - {\delta} { v}_{p,j} \frac{\partial \langle { U}_{p,i} \rangle}{\partial x_j} \, dt   + B_{\delta,ij} \; d{ W}_{\delta,j}  \notag \\ & - \frac{(1+e)(3-e)}{4 \tau_c } \delta v_{p,i} \, dt + \sqrt{\frac{1}{2\tau_c } (1+e)^2 \langle \Theta_p \rangle} \; dW_{c ,i}.
\end{align}
As they represent different physics, the Wiener process for collisions $dW_{c}$ is uncorrelated with $d{ W}_{\delta}$. This collision model is applicable to rapid granular flows that can be modelled by the Boltzmann equation with inelastic hard-sphere collisions, i.e., the collisional and frictional contributions are not accounted for in the particle-phase pressure tensor.

In conclusion, the resulting complete particle stochastic model is the following:
\begin{align}
d \, { x}_{p,i} & = { V}_{p,i} \, dt = ( U_{p,i} + \delta v_{p,i}) \, dt , \\
d \, { U}_{p,i} & = \frac{{ U}_{s,i} - { U}_{p,i}}{\tau_p} \, dt + { g}_i \, dt - \frac{1}{\langle \alpha_p \rangle \rho_p} \frac{\partial \langle \alpha_p \rangle \rho_p \langle P_{ij} \rangle}{\partial x_j} + {\delta} { v}_{p,j} \frac{\partial \langle { U}_{p,i} \rangle}{\partial x_j} \, dt \notag \\ & - \frac{1}{T_{Lp}} ({ U}_{p,i} - \langle { U}_{p,i} \rangle ) \, dt + \sqrt{C_{p} \varepsilon_p} \; d{ W}_{p,i} ,\label{eq:up-final}\\
d \, {\delta} { v}_{p,i} & = - \frac{{\delta} { v}_{p,i}}{\tau_p} \, dt + \frac{1}{\langle \alpha_p \rangle \rho_p} \frac{\partial \langle \alpha_p \rangle \rho_p \langle P_{ij} \rangle}{\partial x_j} - {\delta} { v}_{p,j} \frac{\partial \langle { U}_{p,i} \rangle}{\partial x_j} \, dt  + B_{\delta,ij} \; d{ W}_{\delta,j}  \notag \\ & - \frac{(1+e)(3-e)}{4 \tau_c } \delta v_{p,i} \, dt + \sqrt{\frac{1}{2\tau_c } (1+e)^2 \langle \Theta_p \rangle} \; dW_{c ,i}.
\label{eq:dvp-final}
\end{align}
As in Lagrangian pdf methods for single-phase flows \citep{Pope_turbulent}, it should be borne in mind that such a model is intended to represent the pdf, or the statistical moments, associated with the particle phase, and not the instantaneous particle dynamics, which is ``fictitious''. 

\section{Eulerian model for the fluid phase}
\subsection{Exact Reynolds-average fluid-phase equations}
Taking the RA of \eqref{eq:fluid_cont} yields the transport equation for the RA fluid-phase volume fraction:
\begin{equation}
\frac{\partial \langle \alpha_f \rangle }{\partial t} + \bfnabla \bcdot \langle \alpha_f \rangle \langle {\bf U}_f \rangle_f = 0 .
\label{eq:fluid_cont_RA}
\end{equation}
The PA fluid-phase velocity equation found from \eqref{eq:fluid_NS} is given by
\begin{align}
\frac{\partial \langle \alpha_f \rangle \langle {\bf U}_f \rangle_f}{\partial t} & + \bfnabla \bcdot \langle \alpha_f \rangle ( \langle {\bf U}_f \rangle_f \otimes \langle {\bf U}_f \rangle_f  + \langle {\bf u}_f \otimes {\bf u}_f \rangle _f) = \notag \\ & - \frac{1}{\rho_f} \bfnabla \langle p_f \rangle + \frac{1}{\rho_f} \bfnabla \bcdot \langle \bds{ \sigma}_f \rangle - \langle \alpha_f \rangle \varphi \left\langle \frac{ {\bf U}_f - {\bf U}_p }{\tau_p} \right\rangle_p + \langle \alpha_f \rangle {\bf g}
\label{eq:fluid_NS_RA}
\end{align} 
where 
\begin{equation}\label{eq::mml}
\varphi = \frac{\rho_p \langle \alpha_p \rangle}{\rho_f \langle \alpha_f \rangle}
\end{equation}
is the mean mass loading, and $\langle {\bf u}_f \otimes {\bf u}_f \rangle _f$ is the fluid-phase Reynolds stress tensor.

The transport equation for the fluid-phase Reynolds stress tensor is given by
\begin{align}
& \frac{\partial \langle \alpha_f \rangle \langle {\bf u}_f \otimes {\bf u}_f \rangle_f}{\partial t}  + \nabla \bcdot \langle \alpha_f \rangle ( \langle {\bf U}_f \rangle_f \otimes \langle {\bf u}_f \otimes {\bf u}_f \rangle_f + \langle {\bf u}_f \otimes {\bf u}_f \otimes {\bf u}_f \rangle_f) = \notag \\ & - \langle \alpha_f \rangle (\langle {\bf u}_f \otimes {\bf u}_f \rangle_f \bcdot \nabla \langle {\bf U}_f \rangle _f )^{\dagger} + \frac{1}{\rho_f} ( \nabla \bcdot \langle \bds{\sigma}_f \otimes {\bf u}_f \rangle - \nabla \langle p_f {\bf u}_f \rangle )^{\dagger} \notag \\ & - \frac{1}{\rho_f} (  \langle \bds{\sigma}_f \bcdot \nabla {\bf u}_f \rangle -  \langle p_f \nabla {\bf u}_f \rangle )^{\dagger} \notag \\ & + \frac{\langle \alpha_f \rangle \varphi}{\tau_p} [\langle {\bf u}_f \otimes {\bf u}_p \rangle _p - \langle {\bf u}_f \otimes {\bf u}_f \rangle_p + \langle {\bf u}_f \rangle_p \otimes ( \langle {\bf U}_p \rangle _p - \langle {\bf U}_f \rangle_f )]^{\dagger}.
\label{eq:fluid_Reynolds_RA}
\end{align}
The fluid-phase variables that are averaged with respect to the particle phase, i.e.\ $\langle {\bf u}_f \rangle_p$, appear due to fluid--particle coupling (\eg due to clusters).

\subsection{Modelled RA equations with two-way coupling}
The exact RA fluid-phase equations \eqref{eq:fluid_cont_RA}--\eqref{eq:fluid_Reynolds_RA} could easily be replaced by a suitable RANS model ($k$--$\varepsilon$, Reynolds-stress models, see \cite{Pope_turbulent}); however, single-phase turbulence models typically do not take into account two-way coupling between the phases. Thus, both in the momentum and in the Reynolds-stress equations, we need to formulate the terms that mimic this effect. We follow here the approach first proposed by \cite{Pei_02}.
We consider the direct effect of the particles on the fluid through a random force, since a fluid and a discrete particle will not be present at the same spatial position in the same instant with probability one. 
Thus we define this random force as
\begin{equation}
\mb{A}_{p \rightarrow f} =
\begin{cases}
0  & \text{with a probability} \quad 1- \langle \alpha_p \rangle (t,\mb{x}_f) \\
\bds{\Pi}_{p} & \text{with a probability}\quad \langle \alpha_p \rangle(t,\mb{x}_f) 
\end{cases}
\label{eq:Apf_def}
\end{equation}
where $\bds{\Pi}_{p} $ is a random variable which is
formed from the discrete particles at the location $\mb{x}_p=\mb{x}$
\begin{equation}
\label{def: reverse force}
\bds{\Pi}_{p} \equiv \frac{\rho_p}{\rho_f}
\frac{{\bf U}_{p} - {\bf U}_{s}}{\tau_p}.
\end{equation}
In other words, from the stochastic models for the 
discrete particles, or from the one-point particle pdf value at location 
$\mb{x}=\mb{x}_f$, we form the random variables $\bds{\Pi}_{p}$ with the 
same distribution. This random term mimics the reverse forces due to the 
discrete particles and is only non-zero where the fluid particle is in the
close neighbourhood of a discrete particle. At the location $\mb{x}$
considered, $\bds{\Pi}_{p}$ is defined as a random acceleration term in the 
equation of $\mb{U}_f$, correlated with $\mb{U}_f$, so that we have
\begin{subequations}
\begin{align}
\label{eq:defPi_p:moy}
\lra{ \bds{ \Pi}_{p} } &= \frac{\rho_p}{\rho_f} 
\left\langle \frac{{\bf U}_{p} - {\bf U}_{s}}{\tau_p} \right\rangle = \frac{\rho_p}{\rho_f} 
\left\langle \frac{{\bf U}_{p} - {\bf U}_{f}}{\tau_p} \right\rangle_p, \\
\lra{ \bds{\Pi}_{p}\, \otimes {\bf U}_{f} } &= \frac{\rho_p}{\rho_f} 
\left\langle \frac{\left( {\bf U}_{p} - {\bf U}_{s}\right) \otimes {\bf U}_s}{\tau_p} \right\rangle= \frac{\rho_p}{\rho_f} 
\left\langle \frac{\left( {\bf U}_{p} - {\bf U}_{f}\right) \otimes {\bf U}_f}{\tau_p} \right\rangle_p. \label{eq:defPi_p:corr}
\end{align}
\end{subequations}

Thus, the resulting RA equations for the fluid phase will be
\begin{equation}
\label{eq:feq_alphaf}
\frac{\partial}{\partial t}(\langle \alpha_f \rangle \rho_f) +
\bfnabla \bcdot (\langle \alpha_f \rangle \,\rho_f\lra{ {\bf U}_{f}}) = 0,
\end{equation}
\begin{align}
 \frac{\partial \langle \alpha_f \rangle \langle {\bf U}_{f} \rangle}{\partial t} & + \bfnabla \bcdot \langle \alpha_f \rangle ( \langle {\bf U}_{f} \rangle \otimes \langle {\bf U}_{f} \rangle + \langle {\bf u}_{f} \otimes {\bf u}_{f} \rangle ) = - \frac{1}{\rho_f} \bfnabla \langle p_f \rangle  \notag \\ & +   \nu \nabla^2 \langle {\bf U}_f \rangle + \frac{\langle \alpha_p \rangle \rho_p}{\rho_f} \left\langle \frac{ {\bf U}_{p}  -  {\bf U}_{s} }{\tau_p} \right\rangle + \langle \alpha_f \rangle {\bf g} .
\label{eq:momentum-new}
\end{align}
The resulting Reynolds-stress transport equation will  be composed of the Reynolds stress model, here we use the LRR-IP model~\citep{Pope_turbulent}, and the two-way coupling term of \eqref{eq:defPi_p:corr}:
\begin{align}
& \frac{\partial \langle \alpha_f \rangle \langle {\bf u}_f \otimes {\bf u}_f \rangle}{\partial t}  + \bfnabla \bcdot \langle \alpha_f \rangle ( \langle {\bf U}_f \rangle \otimes \langle {\bf u}_f \otimes {\bf u}_f \rangle) =  \bfnabla \bcdot (\nu   \bfnabla \langle {\bf u}_f \otimes {\bf u}_f \rangle) \notag \\ 
& +{\bds {\mathcal{L}}} + \langle \alpha_f \rangle {\bds {\mathcal{P}}}_f + \langle \alpha_f \rangle {\bds {\mathcal{R}}}_f - \frac{2}{3} \langle \alpha_f \rangle \varepsilon_f {\bf I} + \langle \alpha_p \rangle [ \langle {\bds {\Pi}}_p \otimes {\bf U}_f \rangle - \langle {\bf U}_f \rangle \otimes \langle \bds{\Pi}_p \rangle ]^{\dagger}
\label{eq:reynoldsstress}
\end{align}
where the turbulent mean-gradient production term is defined by 
\begin{equation}
\bds{\mathcal{P}}_f = - (\langle {\bf u}_f \otimes {\bf u}_f \rangle \bcdot \bfnabla \langle {\bf U}_f \rangle )^{\dagger}
\end{equation}
and the pressure-redistribution term is modeled by
\begin{align}
\bds{\mathcal{R}}_f = - C_{Rf} \frac{\varepsilon_f}{k_f} \left( \langle {\bf u}_f \otimes {\bf u}_f \rangle  - \frac{2}{3} k_f {\bf I} \right) 
-C_{2f} \left(  \bds{\mathcal{P}}_f   - \frac{2}{3} {\mathcal{P}}_f  {\bf I} \right)  .
\end{align}
The transport term ${\bds {\mathcal{L}}}$ may be modelled with the different standard models present in literature \citep{Pope_turbulent}. Here we leave it in an unclosed form, since it does not play a role in the homogeneous flows presented in the second part, and also because the transport is exact in the Lagrangian models.

Some remarks are in order concerning the fluid-phase Reynolds-stress model. 
Following \cite{capecelatro2016}, we have chosen the LRR-IP model, which is widely used 
and give reasonably good results, but any other realisable Reynolds-stress model could be chosen, if needed.
The important point is that 
it has been demonstrated that a realisable Reynolds-stress model corresponds to each Lagrangian stochastic model for the fluid \citep{pope1994relationship}. Furthermore,  consistency between Eulerian and Langrangian models of the fluid should be always assured \citep{muradoglu2001hybrid,chibbaro2011note,minier2014guidelines}. 
Notably, the Rotta model is consistent with the standard Langevin model (SLM) for the fluid \citep{Pope_turbulent}, and for this reason it is usually chosen as the standard model to be used in hybrid Eulerian/Lagrangian approach \citep{minier2014guidelines}.

\section{Lagrangian model for the fluid seen by particles}
\label{sec:dilute}
The equation for the particle velocity \eqref{eq:up-final} contains the velocity of the fluid at the position of the particle, or the fluid seen by particles. Since in RANS simulations we have no access to this quantity, nor to its average, a model for it has to be specified.
Furthermore, given that we are considering flow with two-way coupling, this effect has to be included also in this fluid-phase equation.
The model for ${\bf U}_s$ will be a Langevin equation of the type
\begin{equation}
d U_{s,i} = [ A_{s,i} + A_{p \rightarrow s,i} ] \, dt + B_{s,ij} \, dW_{s,j} 
\end{equation}
where $A_{p \rightarrow s,i}$ represents explicitly the effect of the particles on the fluid, and $dW_{s,j}$ is a different Wiener process with respect to those present in equations \eqref{eq:up-final} and \eqref{eq:dvp-final}. 
The drift term, ${\bf A}_s$, is modeled as done for dilute flows \citep{Min_04,Pei_06}:
\begin{equation}
{ A}_{s,i} = - \frac{1}{\rho_f} \frac{\partial \langle p_f \rangle}{\partial x_i} + (\langle { U}_{p,j} \rangle - \langle { U}_{f,j} \rangle ) \frac{\partial \langle { U}_{f,i} \rangle}{\partial x_j} 
+G_{ij}\left( { U}_{s,j}-\lra{{ U}_{f,j}} \right)
+ { g} _i
\label{eq:drift2}
\end{equation}
where 
\begin{equation}\label{matriceG}
G_{ij}=-\frac{1}{{ T}_{L,i}^*}\delta_{ij}+G_{ij}^a
\end{equation}
is the matrix defining the corresponding Reynolds stress model.
The first part is the simplified Langevin model (SLM) adapted to the inertial particles,
while ${\bf G}^a$ is a traceless matrix to be added to generalise the model.
For instance, the LRR-IP model reads
\begin{equation}
G_{ij}^a= C_{2f} \frac{\partial \langle { U}_{f,i} \rangle}{\partial x_j}  
\end{equation}
 and $C_{2f}$ is the IP constant, usually taken $C_{2f}=\frac{3}{5}$
consistent with rapid-distorsion theory. 

The crossing trajectory effect (CTE) has been modelled in \eqref{matriceG} by using the timescale according to Csanady's analysis:
\begin{equation}
T_{L,i}^* = \frac{T_{Lf}}{\sqrt{1 + \zeta_i \beta^2 \frac{3 |\langle {\bf U}_r \rangle |^2}{2k_f}}}
\end{equation}
where
\begin{equation}
T_{Lf} = \frac{1}{\left( \frac{1}{2} + \frac{3}{4} C_{0f} \right)} \frac{k_f}{\varepsilon_f}
\end{equation}
is the Lagrangian time scale, $C_{0f}$ being linked to the Rotta constant by the relation
\begin{equation}
C_{0f} = \frac{2}{3} \Bigl ( C_R - 1 + C_{2f} \frac{\mathcal{P}}{\varepsilon_f} \Bigl ) ,
\end{equation}
and the relative velocity is defined by
\begin{equation}
{\bf U}_r = {\bf U}_p - {\bf U}_s.
\end{equation}
Moreover, $\zeta_1 =1$ in the mean drift direction and $\zeta_{2,3} =4$ in the cross directions, $\beta = T_{Lf} / T_{Ef}$ is the ratio of the Lagrangian and the Eulerian timescales \citep{wang1993dispersion}.

The modelling of the two-way coupling term instead is as follows.
The exact expression for the two-way coupling term,
${\bf A}_{p \rightarrow s}$, which is induced by
the presence of the discrete particles, is not \emph{a priori} known. The
underlying force corresponds to the exchange of momentum between the
fluid and the particles, but should not be confused with the total
force acting on particles since the latter includes external forces
such as gravity. The effect of particles on fluid properties is
expressed directly in the stochastic equation of ${\bf U}_s$ through a
simple stochastic model. The force exerted by one
particle on the fluid corresponds to the drag force written here as
\begin{equation}
\label{drag term}
{\bf F}_{p \rightarrow f}= - m_p \frac{{\bf U}_s-{\bf U}_p}{\tau_p}
\end{equation}
where $m_p$ is the mass of a particle. The total force acting on the
fluid element surrounding a discrete particle  is then obtained as the
sum of all elementary forces, ${\mb {F}}_{p \rightarrow f}$, and the
resulting acceleration is modelled here as \citep{Pei_02}
\begin{equation}
{ A}_{p \rightarrow s,i} = 
-\varphi \frac{{ U}_{s,i} - { U}_{p,i}}{\tau_p}
\label{eq:drift3}
\end{equation}
where $\varphi$ is the mean mass loading introduced in \eqref{eq::mml}.

\section{Closure of the diffusion coefficient}
The drift terms are given by \eqref{eq:drift2} and \eqref{eq:drift3}, but the diffusion coefficient needs to be specified in order to obtain a proper closure. Analogously to dilute flows, we look for a diffusion matrix in a diagonal, but anisotropic form.
To close this term we consider the decay of the turbulent kinetic energy in the homogeneous case, in absence of mean shear, and we make the following assumption
\begin{equation}
\frac{d k_f}{dt} \simeq \frac{d k_{f@p}}{dt}
\label{eq:costraint-us}
\end{equation}
where $k_f = \frac{1}{2}  \langle {\bf u}_f \bds{\bcdot} {\bf u}_f \rangle$ is the turbulent kinetic energy of the fluid phase and $k_{f@p} =  \frac{1}{2} \langle {\bf u}_f \bds{\bcdot} {\bf u}_f \rangle_p$ is the turbulent kinetic energy of the fluid phase seen by the particles. A possible alternative could be to impose an analogous relation, but for all the Reynolds-stress components, which would lead to a much more complex model.

Recalling the decomposition of the fluid velocity shown in \eqref{eq:decomp}, we can obtain the following equality
\begin{equation}
\langle {\bf u}_f \rangle_p = \langle {\bf U}_f - \langle {\bf U}_f \rangle_f \rangle_p = \langle {\bf U}_s \rangle - \langle {\bf U}_f \rangle.
\end{equation}
Thus, \eqref{eq:costraint-us} can be rewritten as
\begin{equation}
\frac{d k_f}{dt} = \frac{1}{2} \frac{d}{dt} \sum_{i=1}^3 [  \langle { U}_{s,i}^2 \rangle + \langle { U}_{f,i} \rangle^2 - 2\langle { U}_{s,i} \rangle \langle { U}_{f,i} \rangle ] 
\label{eq:closure}
\end{equation}
where the fluid velocity and the fluid velocity seen temporal variation in the homogeneous case are expressed by
\begin{flalign}
 &\dfrac{d \langle { U}_{f,i} \rangle}{dt}  =  f_i + {g}_i + \varphi \dfrac{\langle { U}_{p,i} - { U}_{s,i} \rangle}{\tau_p} \, , \label{eq:closure1}\\
 & d { U}_{s,i}  =  \langle \alpha_f \rangle f_i \, dt - \frac{1}{{ T}_{L,i}^*} ({ U}_{s,i} - \langle { U}_{f,i} \rangle ) \, dt + { g}_i \, dt + \varphi \frac{ { U}_{p,i} - { U}_{s,i} }{\tau_p}\, dt + { B}_{s,ii} \, d { W}_{s,i} \,, \label{eq:closure2}\\
& \frac{d \langle { U}_{s,i}^2 \rangle}{dt} = 2  \langle \alpha_f \rangle f_i \langle U_{s,i} \rangle - \frac{2}{{ T}_{L,i}^*}  \langle { U}_{s,i}^2 \rangle + \frac{2}{{ T}_{L,i}^*} \langle { U}_{s,i} \rangle \langle { U}_{f,i} \rangle + 2 { g}_i  \langle { U}_{s,i} \rangle \notag \\ & \qquad \qquad + 2 \varphi \frac{\langle { U}_{p,i} { U}_{s,i} \rangle - \langle { U}_{s,i}^2  \rangle}{\tau_p} + { B}_{s,ii}^2 \, , \label{eq:closure3} \\
 & \frac{d \langle { U}_{f,i} \rangle^2 }{dt} = 2 \langle { U}_{f,i} \rangle  \frac{d \langle { U}_{f,i} \rangle}{dt} = 2 f_i \langle U_{f,i} \rangle + 2 { g}_i  \langle { U}_{f,i} \rangle + 2 \varphi \langle { U}_{f,i} \rangle  \frac{\langle { U}_{p,i} - { U}_{s,i} \rangle}{\tau_p} \,, \label{eq:closure4}\\
& \frac{d \langle { U}_{s,i} \rangle  \langle { U}_{f,i} \rangle}{dt} = - \frac{1}{{ T}_{L,i}^*} (\langle { U}_{f,i} \rangle \langle { U}_{s,i} \rangle - \langle { U}_{f,i} \rangle^2 ) + { f}_i  ( \langle \alpha_f \rangle \langle { U}_{f,i} \rangle  +  \langle { U}_{s,i} \rangle)  \notag \\ &\qquad \qquad  \, \, + { g}_i  ( \langle { U}_{f,i} \rangle +  \langle { U}_{s,i} \rangle) + \varphi \langle { U}_{f,i} \rangle  \frac{\langle { U}_{p,i} - { U}_{s,i} \rangle}{\tau_p}    + \varphi \langle { U}_{s,i} \rangle \frac{\langle { U}_{p,i} - { U}_{s,i} \rangle}{\tau_p} \, .
\label{eq:closure5}
\end{flalign}

with $f_i = -1 / (\langle \alpha_f \rangle \rho_f ) \partial \langle p_f \rangle / \partial x_i$. Now, if we substitute \eqref{eq:closure3}--\eqref{eq:closure5} into \eqref{eq:closure}, the following relation is obtained:
\begin{align}
& \frac{d k_f}{dt}  = -  \varepsilon_f  + \sum_{i=1}^3 \left[ \varphi \frac{\langle U_{p,i} U_{s,i} \rangle - \langle U_{s,i}^2 \rangle}{\tau_p} - \varphi \langle U_{f,i} \rangle \frac{\langle U_{p,i} - U_{s,i} \rangle}{\tau_p} \right] \notag \\ &  = \sum_{i=1}^3 \left[ - \frac{1}{T_{L,i}^*} \langle ( U_{s,i}  - \langle U_{f,i} \rangle )^2 \rangle - \langle \alpha_p \rangle f_i (\langle U_{s,i} \rangle - \langle U_{f,i} \rangle) + \frac{B_{s,ii}^2}{2} + \varphi \frac{\langle u_{p,i} u_{s,i} \rangle - \langle u_{s,i}^2 \rangle}{\tau_p}  \right] \, .
\end{align}
The terms $(1/{ T}_{L,i}^* ) \langle ( U_{s,i}  - \langle U_{f,i} \rangle )^2 \rangle $ and $\varepsilon_f$ can be rearranged together, whereas the other terms are redistributed on each corresponding component
shown above. 
The result is 
\begin{multline}
{B}_{s,ii}^2 =  2 \varphi  \frac{\langle { U}_{p,i} - { U}_{s,i} \rangle}{\tau_p}  (\langle { U}_{s,i} \rangle - \langle { U}_{f,i} \rangle) + 2 \langle \alpha_p \rangle f_i (\langle U_{s,i} \rangle - \langle U_{f,i} \rangle)    \\  
+ \varepsilon_f \left[ C_0 { b}_i \frac{\tilde{k}_f}{k_f} + \frac{2}{3} \left( { b}_i \frac{\tilde{k}_f}{k_f} -1 \right) \right] 
\label{eq:diff_coeff}
\end{multline}
with ${b}_i = T_L / { T}_{L,i}^*$ and
\begin{equation}
\tilde{k}_f = \frac{3}{2} \frac{ \sum_{i=1}^3 { b}_i \langle ( U_{s,i}  - \langle U_{f,i} \rangle )^2 \rangle}{\sum_{i=1}^3{ b}_i } \, .
\end{equation}

Having pointed out all the relevant terms, we can now rewrite the set of stochastic equations with all the terms shown explicitly:
%
\begin{align} 
& d{ x}_{p,i}(t)= ({ U}_{p,i} + \delta v_{p,i} )\,dt \, , \label{eq:SDEp-finalx}\\
& d \, { U}_{p,i}(t) = \frac{{ U}_{s,i} - { U}_{p,i}}{\tau_p} \, dt + { g}_i \, dt - \frac{1}{\langle \alpha_p \rangle \rho_p} \frac{\partial \langle \alpha_p \rangle \rho_p \langle P_{ij} \rangle}{\partial x_j} + {\delta} { v}_{p,j} \frac{\partial \langle { U}_{p,i} \rangle}{\partial x_j} \, dt \notag \\ 
& \qquad \qquad - \frac{1}{T_{Lp}} ({ U}_{p,i} - \langle { U}_{p,i} \rangle ) \, dt + \sqrt{C_{p} \varepsilon_p} \; d{ W}_{p,i} ,\label{eq:SDEp-finalup}\\
& d \, {\delta} { v}_{p,i}(t) = - \frac{{\delta} { v}_{p,i}}{\tau_p} \, dt + \frac{1}{\langle \alpha_p \rangle \rho_p} \frac{\partial \langle \alpha_p \rangle \rho_p \langle P_{ij} \rangle}{\partial x_j} - {\delta} { v}_{p,j} \frac{\partial \langle { U}_{p,i} \rangle}{\partial x_j} \, dt + B_{\delta,ij} \; d{ W}_{\delta,j} \notag \\ 
& \qquad \qquad   - \frac{(1+e)(3-e)}{4 \tau_c } \delta v_{p,i} \, dt + \sqrt{\frac{1}{2\tau_c } (1+e)^2 \langle \Theta_p \rangle} \; dW_{c ,i}
\label{eq:SDEp-finaldelta} \\
& d{ U}_{s,i}(t) = - \frac{1}{\rho_f} \frac{\partial \langle p_f \rangle}{\partial x_i} \, dt + (\langle { U}_{p,j} \rangle - \langle { U}_{f,j} \rangle )  \frac{\partial \langle { U}_{f,i} \rangle}{\partial x_j} \, dt \notag\\ 
& \qquad \qquad + G_{ij}({ U}_{s,j} - \langle { U}_{f,j} \rangle )  \,dt  - \varphi \left( \frac{{ U}_{s,i} - { U}_{p,i} }{\tau_p} \right)\, dt   + { g}_i \, dt \notag\\ 
& \qquad \qquad + \Bigl [ \varepsilon_f \left( C_0 {b}_i \frac{\tilde{k}_f}{k_f} + \frac{2}{3} \left( {b}_i \frac{\tilde{k}_f}{k_f} -1 \right) \right)  +  2 \varphi  \frac{\langle { U}_{p,i} - { U}_{s,i} \rangle}{\tau_p} (\langle { U}_{s,i} \rangle - \langle { U}_{f,i} \rangle) \notag\\ 
& \qquad \qquad \qquad - 2 \frac{\langle \alpha_p \rangle}{\langle \alpha_f \rangle \rho_f} \frac{\partial \langle p_f \rangle}{\partial x_i}  (\langle { U}_{s,i} \rangle - \langle { U}_{f,i} \rangle  ) \Bigl ]^{1/2}  \,  \, d{ W}_{s,i} \, . \label{eq:SDEp-finalus}
\end{align}
The set of equations for the particle and fluid velocities can be recast in vector form as follows:
\begin{equation}
d \, {\bf X} = {\bf A } \, dt + [B]  d{\bf W} \label{eq:system_upus} 
\end{equation}
where ${\bf A}$ is the drift term, and
\begin{equation}
{\bf X} = \left(
\begin{array}{c}
{\bf U}_{p}\\
\delta {\bf v}_{p} \\
{\bf U}_{s}\\
\end{array}
\right) \qquad [ B] = \left[
\begin{array}{cccc}
C [I] & 0 & 0 & 0\\
0 & [B_{\delta}] & K[I] & 0\\
0 & 0 & 0 & [B_s] \\
\end{array}
\right] \qquad d {\bf W} = \left(
\begin{array}{c}
d {\bf W}_{p}\\
d {\bf W}_{\delta}\\
d {\bf W}_{c} \\
d {\bf W}_{s}\\
\end{array}
\right)
\end{equation}
$C = \sqrt{C_p \varepsilon_p}$ is the diagonal diffusion coefficient in the equation of the correlated velocity and $K = \sqrt{1/(2 \tau_c) (1+e^2) \langle \Theta_p \rangle}$ is the diagonal diffusion coefficient for the collisions in the uncorrelated velocity equation.

\section{RA macroscopic equations}
\label{sec:mean_eq_p}
The state vector of the Lagrangian description, given by ${\bf Z} = ( {\bf x}_p, {\bf U}_p, \bds{\delta} {\bf v}_p, {\bf U}_s)$, is associated with a single particle, while $\langle {\bf Z} \rangle$ stands for $\langle {\bf Z} \rangle [{\bf x}^{(k)}]$. The particle system is thus represented by this set of Lagrangian equations, where the particle state variables are modeled through a Langevin equation, or to be more rigorous as a diffusion process.
This set of Lagrangian stochastic equations for the trajectories of the sample particles corresponds, from the pdf point of view, to the following Fokker--Planck (FP) equation for the Eulerian mass density function (mdf) \citep{Pop_85,Min_01,Fox2003}:
\begin{align}
\label{eq:FP_pL_fp}
& \frac{\partial F^{E}_{p}}{\partial t}  + 
\mathrm{U}_{p,i}\frac{\partial F^{E}_{p}}{\partial x_{i}} + 
\delta \mathrm{v}_{p,i}\frac{\partial F^{E}_{p}}{\partial x_{i}} = \notag \\ 
- &
\frac{\partial}{\partial \mathrm{U}_{p,i}}
                  \left( \left[\frac{(\mathrm{U}_{s,i} - \mathrm{U}_{p,i})}{\tau_p} - \frac{1}{\langle \alpha_p \rangle \rho_p} \frac{\partial \langle \alpha_p \rangle \rho_p \langle P_{ij} \rangle}{\partial x_j} + \delta \mathrm{v}_{p,j} \frac{\partial \langle U_{p,i}\rangle}{\partial x_j} -  \frac{1}{T_{Lp}} ( \mathrm{U}_{p,i} - \langle U_{p,i} \rangle) + g_i \right]  F^{E}_{p} \right) \notag \\ 
 & + \frac{1}{2} C_p \varepsilon_p  \delta_{ij} \frac{\partial ^2 F_p^E}{\partial \mathrm{U}_{p,i} \partial \mathrm{U}_{p,j}}
 \notag\\
 -&\frac{\partial}{\partial \delta \mathrm{v}_{p,i}} 
                  \left( \left[-\frac{\delta \mathrm{v}_{p,i}}{\tau_p} + \frac{1}{\langle \alpha_p \rangle \rho_p} \frac{\partial \langle \alpha_p \rangle \rho_p \langle P_{ij} \rangle}{\partial x_j} - \delta \mathrm{v}_{p,j} \frac{\partial \langle U_{p,i}\rangle}{\partial x_j}   - \frac{(1+e)(3-e)}{4\tau_c} \delta \mathrm{v}_{p,i} \right] F^{E}_{p} \right) \notag \\ 
 & + \frac{1}{2} \left[({ B}_{\delta} { B}_{\delta}^T)_{ij} + \frac{1}{2 \tau_c} (1+e)^2 \langle \Theta_p \rangle \delta_{ij} \right]  \frac{\partial ^2 F_p^E}{\partial \delta \mathrm{v}_{p,i} \partial \delta \mathrm{v}_{p,j}} \notag \\ 
 &
 + \left[\frac{1}{\rho} \frac{\partial \langle p_f \rangle}{\partial x_i}  - (\langle U_{p,j} \rangle - \langle U_{f,j} \rangle) \frac{\partial \langle U_{f,i} \rangle}{\partial x_j}  -g_i \right] \frac{\partial F_p^E}{\partial \mathrm{U}_{s,i}} \notag \\ 
 &- \frac{\partial}{\partial \mathrm{U}_{s,i}} \left[ \left( G_{ij}(\mathrm{U}_{s,j} - \langle U_{f,j}  \rangle) - \varphi \frac{(\mathrm{U}_{s,i} - \mathrm{U}_{p,i})}{\tau_p} \right) F_p^E \right] + \frac{1}{2} B_{s,ii}^2 \frac{\partial^2 F_p^E}{\partial \mathrm{U}_{s,i} \partial \mathrm{U}_{s,i}}
\end{align}
where $B_{s,ii}$  is the diffusion matrix given by \eqref{eq:diff_coeff}, and it is not given here explicitely for the sake of clarity. 
$F_p^E(t, \mathbf{x}; \bds{\mathrm{U}}_p, \bds{\delta}\bds{\mathrm{v}}_p, \bds{\mathrm{U}}_s )$ is the probable mass of discrete particles in an element in the phase-space of volume $d\bds{\mathrm{U}}_p \, d\bds{\delta}\bds{\mathrm{v}}_p \,d\bds{\mathrm{U}}_s$ at a position $\mathbf{x}$.
The FP equation can be used to derive the PA equations for the particle phase using the definition
\begin{equation}
\langle \alpha_p \rangle (t,\mb{x})\rho_p\lra{\mathbb{O}}_p (t,\mb{x}) :=
\int \mathbb{O}(\bds{\mathrm{U}}_p, \bds{\delta}\bds{\mathrm{v}}_p, \bds{\mathrm{U}}_s) \;F^E_p(t,\mb{x};\bds{\mathrm{U}}_p, \bds{\delta}\bds{\mathrm{v}}_p, \bds{\mathrm{U}}_s)
\,d\bds{\mathrm{U}}_p\, d\bds{\delta}\bds{\mathrm{v}}_p \,d\bds{\mathrm{U}}_s
\label{eq:def_mean}
\end{equation}
where $\mathbb{O}$ is a generic observable attached to a discrete particle.
It is important to note that, there is a perfect equivalence between PA Eulerian quantities and particle-average Lagrangian quantities. Since we employ a pdf approach with a trajectories point of view, i.e.\ Lagrangian, we prefer to adopt in the rest of the work the particle-average notation, remarking that
\begin{equation}
\langle {\bf U}_f \rangle = \langle {\bf U}_f \rangle_f, \qquad \; \; \langle {\bf U}_p \rangle = \langle {\bf U}_p \rangle_p, \qquad \; \; \langle {\bf U}_s \rangle = \langle {\bf U}_f \rangle_p,
\label{eq:analogy}
\end{equation}
and analogously for second-order moments.

Closed RA transport equations can now be derived either from the FP or Lagrangian equations. 
The so-obtained RA continuity equation is the following
\begin{equation}
\label{eq:feq_alphap}
\frac{\partial}{\partial t}(\langle \alpha_p \rangle \rho_p) +
\bfnabla \bcdot (\langle \alpha_p\rangle \,\rho_p\lra{ {\bf U}_{p}}) = 0.
\end{equation}
The momentum equation reads
\begin{equation}
\label{eq:feq_Up}
\langle \alpha_p \rangle \,\rho_p\frac{D}{Dt}\lra{{\bf U}_{p}} =
- \bfnabla \bcdot \langle \alpha_p \rangle \,\rho_p (\lra{ {\bf u}_{p} \otimes {\bf u}_{p}} + \langle {\bf P} \rangle)
+\langle \alpha_p \rangle \,\rho_p \left\langle \frac{{\bf U}_s - {\bf U}_p}{\tau_p} \right\rangle \,+\, \langle \alpha_p \rangle {\bf g} .
\end{equation}
where $D/Dt = \partial/\partial t + \lra{{\bf U}_{p}} \bcdot \bfnabla$.
Second-order moments, on the other hand, give the following equations for the particle-phase Reynolds stress: 
\begin{align}
\label{eq:feq_part-rey}
\langle \alpha_p \rangle \rho_p\frac{D}{Dt}\lra{{\bf u}_{p} \otimes {\bf u}_{p}}  & =
- \bfnabla \bcdot
(\langle \alpha_p \rangle \,\rho_p\lra{{\bf u}_{p} \otimes {\bf u}_{p} \otimes {\bf u}_{p}})
-\langle \alpha_p \rangle \rho_p (\lra{{\bf u}_{p} \otimes {\bf u}_{p}} \bcdot \bfnabla \lra{{\bf U}_{p}})^{\dagger} \notag \\
& +\frac{\langle \alpha_p \rangle \,\rho_p}{\tau_p} ( \lra{ {\bf u}_{s} \otimes {\bf u}_{p} } - \langle {\bf u}_p \otimes {\bf u}_p \rangle ) ^{\dagger} + \langle \alpha_p \rangle \rho_p \bds{\mathcal{R}}_p - \langle \alpha_p \rangle \rho_p \bds{\varepsilon}_p 
\end{align}
where the redistribution is expressed by
\begin{align}
\bds{\mathcal{R}}_p = - C_{Rp} \frac{\varepsilon_p}{k_p} \left( \langle {\bf u}_p \otimes {\bf u}_p \rangle  - \frac{2}{3} k_p {\bf I} \right) 
\end{align}
with
\begin{equation}
C_{Rp} = 1 + \frac{3}{2} C_{0p} ,
\end{equation}
and the dissipation tensor is closed using the Rotta model \citep{Pope_turbulent}:
\begin{align}\label{eq:::ep}
\bds{\varepsilon}_p =  \varepsilon_p \left[ f_s \frac{\langle {\bf u}_p \otimes {\bf u}_p \rangle}{k_p} + (1-f_s) \frac{2}{3} {\bf I} \right] .
\end{align}
The transport equation for the particle-phase pressure tensor is
\begin{align}
\label{eq:feq_part-gran}
\langle \alpha_p \rangle \rho_p\frac{D}{Dt}\lra{{\bf P} }  =
& - \bfnabla \bcdot
[\langle \alpha_p \rangle \,\rho_p (\lra{{\bf u}_{p} \otimes {\bf P}} + \langle {\bf Q}\rangle)]
-\langle \alpha_p \rangle \rho_p (\lra{{\bf P}} \bcdot \bfnabla \lra{{\bf U}_{p}})^{\dagger}  + \langle \alpha_p \rangle \rho_p \bds{\varepsilon}_p \notag \\ & - \frac{2}{\tau_p}\langle \alpha_p \rangle \rho_p \langle {\bf P}  \rangle + \frac{1}{2 \tau_c} [(1+e)^2 \langle \Theta_p \rangle {\bf I} - (1+e)(3-e) \langle {\bf P} \rangle ] .
\end{align}
Comparing \eqref{eq:feq_part-rey}, \eqref{eq:feq_part-gran} to \eqref{eq:pressure_RA}, \eqref{eq:particle_reynolds_RA}, we can see that the closed terms have been reproduced correctly in the Lagrangian model, while the previously unclosed terms such as dissipation are now modelled.

Particular attention should be given to the closure of the term $\bds{\varepsilon}_{p} =\langle {\bf P} \bcdot \nabla {\bf u}_p \rangle$, which plays the role of a sink in the equation of the particle-phase turbulent kinetic energy and of a source in the equation of the particle-phase pressure tensor. In analogy to single-phase flow, where dissipation of turbulent kinetic energy leads to viscous heating, it is modelled as a particle-phase anisotropic dissipation tensor $\bds{\varepsilon}_p$, whose trace divided by two gives the scalar particle-phase dissipation $\varepsilon_p$.  As shown in \cite{Fox2015,capecelatro2016strongly}, when the mean mass loading is significant, the particle-phase pressure tensor is highly anisotropic due to the source term $\bds{\varepsilon}_p$ (i.e., $f_s \approx 0.93$ in \eqref{eq:::ep}). 

The transport equation of the scalar particle-phase dissipation is modelled as follows:
\begin{multline}
 \frac{\partial \langle \alpha_p \rangle \varepsilon_p }{\partial t} + \bfnabla \bcdot ( \langle \alpha_p \rangle \langle {\bf U}_{p} \rangle \varepsilon_p ) 
 = \bfnabla \bcdot \left[ \langle \alpha_p \rangle \left( \nu_p + \frac{\nu_{p,t}}{\sigma_{\epsilon,p}} \right) \bfnabla \varepsilon_p \right] \\ 
 - \langle \alpha_p \rangle C_{\epsilon 1 p} \langle {\bf u}_p \otimes {\bf u}_p \rangle \bds{:} \bfnabla \langle {\bf U}_p \rangle \frac{\varepsilon_p}{k_p}  - \langle \alpha_p \rangle C_{\epsilon 2 p} \frac{\varepsilon_p ^2}{k_p} + \langle \alpha_p \rangle \frac{C_{3p}}{\tau_p} \left( \frac{ k_{fp}}{k_{f@p} } \varepsilon_f - \beta_p \, \varepsilon_p \right) ,
 \label{eq:epsilon_p}
\end{multline}
which differs slightly from the model in \cite{Fox2014} because $k_{fp}$ is known in the Lagrangian model proposed here.
Now that we have modelled the particle-phase dissipation, we can define a timescale for the particle phase $T_p = k_p / \varepsilon_p $ to be used in the model equation for the fluid-phase dissipation.  

For the fluid-phase dissipation, we propose the following model, built with the standard single-phase fluid dissipation model equation \citep{Pope_turbulent} and an additional contribution due to particle-fluid interaction:
\begin{multline}
 \frac{\partial \langle \alpha_f \rangle \varepsilon_f }{\partial t} + \bfnabla \bcdot ( \langle \alpha_f \rangle \langle {\bf U}_{f} \rangle \varepsilon_f ) 
 = \bfnabla \bcdot \left[ \langle \alpha_f \rangle \left( \nu + \frac{\nu_t}{\sigma_{\epsilon}} \right) \bfnabla \varepsilon_f \right]  \\
 - \langle \alpha_f \rangle C_{\epsilon 1 f} \langle {\bf u}_f \otimes {\bf u}_f \rangle \bds{:} ( \bfnabla \langle {\bf U}_f \rangle ) \frac{\varepsilon_f}{k_f}   
 - \langle \alpha_f \rangle C_{\epsilon 2 f} \frac{\varepsilon_f ^2}{k_f}  \\
 +  \frac{ \rho_p \langle \alpha_p \rangle}{\rho_f} \frac{C_{3 f}}{\tau_p} \left(  \frac{ k_{fp}}{k_{f@p} } \varepsilon_p - \beta_f \varepsilon_f \right)    
 + \frac{ \rho_p \langle \alpha_p \rangle}{\rho_f} \frac{C_4}{\tau_p} \frac{(\langle {\bf U}_p \rangle - \langle {\bf U}_f \rangle) \bcdot \langle {\bf u}_d \rangle }{2}  \frac{\varepsilon_p}{k_p} 
 \label{eq:epsilon-new}
\end{multline}
where $\langle {\bf u}_d \rangle = \langle {\bf U}_s \rangle - \langle {\bf U}_f \rangle$. Here, we have split the total energy rate dissipation into two contributions, deriving from the energy exchange between phases (fourth term on the r.h.s.), and from the drag production (last term).

The Lagrangian approach is tantamount to computing the entire pdf of the variables considered in the state vector. With respect to an Eulerian moment approach, it means that more information is available. 
Notably, we wish to derive here the RA equations for  the mean fluid velocity seen by the particles $\lra{{\bf U}_{s}}$, and for all the second-order  velocity moments, 
$\lra{{\bf u}_{s} \otimes {\bf u}_{s}}$, $\lra{{\bf u}_{s} \otimes {\bf u}_{p}}$. 
We can obtain the RA equations starting from the transport equation of the Eulerian mdf $F_p^E$ \eqref{eq:FP_pL_fp}:
\begin{multline}
\label{eq:feq_Us}
\langle \alpha_p \rangle \,\rho_p\frac{D}{Dt}\lra{{\bf U}_{s}}  = 
-\bfnabla \bcdot (\langle \alpha_p \rangle \,\rho_p\lra{{\bf u}_{s} \otimes  {\bf u}_{p}}) \\
+ \langle \alpha_p \rangle\,\rho_p \left[ - \frac{1}{\rho_f} \bfnabla \langle p_f \rangle 
+ ( \bfnabla \langle {\bf U}_f \rangle ) \bcdot (\langle {\bf U}_p \rangle - \langle {\bf U}_f \rangle ) + 
{\bf G} \bcdot (\langle {\bf U}_s \rangle - \langle {\bf U}_f \rangle) 
+ {\bf g} \right] \\
-\langle \alpha_p \rangle\,\rho_p \varphi \left( \frac{\langle {\bf U}_s -{\bf U}_p \rangle}{\tau_p} \right)
\end{multline}
where $D/Dt = \partial/\partial t + \lra{{\bf U}_{p}} \bcdot \bfnabla$. 

For the second-order moments we obtain 
\begin{multline}
\label{eq:feq_upus}
\langle \alpha_p \rangle  \rho_p  \frac{D}{Dt}\lra{{\bf u}_{s} \otimes {\bf u}_{p}}  = -\bfnabla \bcdot [ \langle \alpha_p \rangle\,\rho_p (
 \lra{{\bf u}_{s} \otimes {\bf u}_{p} \otimes {\bf u}_{p}} + \langle {\bf u}_{s} \otimes \bds{\delta}{\bf v}_{p} \otimes \bds{\delta}{\bf v}_{p} \rangle)]  \\ 
-\langle \alpha_p \rangle \rho_p (\lra{{\bf u}_{s} \otimes {\bf u}_{p}} \bcdot \bfnabla \lra{{\bf U}_{p}}^{T}) -\langle \alpha_p \rangle \rho_p [(\lra{{\bf u}_{p} \otimes {\bf u}_{p}} + \langle \bds{\delta}{\bf v}_{p} \otimes \bds{\delta}{\bf v}_{p} \rangle)  \bcdot \bfnabla \lra{{\bf U}_{s}}^{T}] \\  
+\langle \alpha_p \rangle \,\rho_p 
{\bf G}\bcdot \lra{ {\bf u}_{s} \otimes {\bf u}_{p} }^{T} 
+\langle \alpha_p \rangle\,\rho_p \varphi \frac{\langle {\bf u}_p \otimes {\bf u}_p \rangle - \langle {\bf u}_s \otimes {\bf u}_p \rangle}{\tau_p} \\ 
-\langle \alpha_p \rangle \,\rho_p \frac{1}{{\bf T}_{Lp}} \circ \lra{ {\bf u}_{p} \otimes {\bf u}_{s} } +\langle \alpha_p \rangle \,\rho_p \frac{\langle {\bf u}_s \otimes {\bf u}_s \rangle - \langle {\bf u}_p \otimes {\bf u}_s \rangle}{\tau_p} 
\end{multline}
and
\begin{multline}
\label{eq:feq_usus}
\langle \alpha_p \rangle \rho_p\frac{D}{Dt}\lra{{\bf u}_{s} \otimes {\bf u}_{s}}  = 
- \bfnabla \bcdot
(\langle \alpha_p \rangle \,\rho_p\lra{{\bf u}_{s} \otimes {\bf u}_{s} \otimes {\bf u}_{s}})
-\langle \alpha_p \rangle \rho_p (\lra{{\bf u}_{s} \otimes {\bf u}_{s}} \bcdot \bfnabla \lra{{\bf U}_{s}})^{\dagger}  \\
+\langle \alpha_p \rangle \,\rho_p 
({\bf G} \bcdot \lra{ {\bf u}_{s} \otimes {\bf u}_{s} })
^{\dagger} 
+
\langle \alpha_p \rangle\,\rho_p \varphi 
\frac{  \langle {\bf u}_s \otimes {\bf u}_p \rangle^{\dagger} - 2 \langle {\bf u}_s \otimes {\bf u}_s \rangle}{\tau_p}  
+ \langle \alpha_p \rangle \,\rho_p\lra{{\bf B}_s {\bf B}_s^T}
\end{multline}
where the $\circ$ symbol denotes an element-by-element product.

Eulerian transport equations for the cross-correlations $\langle {\bf u}_p \otimes \delta {\bf v}_p \rangle$ and $\langle {\bf u}_s \otimes \delta {\bf v}_p \rangle$ could also be written to demonstrate that $\delta {\bf v}_p$ is uncorrelated with respect to the other variables. Also, it is important to note that these moment equations are, in general, not closed (\eg the turbulent fluxes involve the third-order moments). However, for statistically homogeneous flows such as particle-laden isotropic turbulence \citep{fevrier2005partitioning,sundaram1999numerical,elghobashi1994predicting} and Cluster-Induced-Turbulence (CIT) \citep{Fox2015}, the spatial gradients are zero, and a closed set of moment equations results.  

\section{Discussion and conclusions}

The main objective of this work was to develop a Lagrangian pdf model for particle-laden turbulent flows valid for all mean mass loadings (yet more accurately with $\alpha_p < 0.2$) and including particle-particle collisions.  In order to correctly account for the latter, the particle-phase kinetic energy must be decomposed into two components, namely, the spatially correlated and uncorrelated contributions. In the Lagrangian pdf model, this decomposition is taken into account by introducing two separate particle-phase velocity variables ($\mathbf{U}_p$ and $\delta \mathbf{v}_p$), which are statistically uncorrelated. Another important feature of the Lagrangian pdf model is the distinction between the fluid-phase velocity $\mathbf{U}_f$ and the fluid velocity seen by the particles $\mathbf{U}_s$. When the mean mass loading is non-negligible, the dynamics of the fluid velocity seen by the particles is strongly affected by coupling with the particle phase.  The Lagrangian pdf model thus provides a closure for the moments of $\mathbf{U}_s$, which are not available from the macroscopic Reynolds-average equations.

We have proposed a stochastic model which represents the joint state variables ($\mathbf{x}_p,\mathbf{V}_p,\delta \mathbf{v}_p,\mathbf{U}_s$) as a diffusion process, or informally a Langevin equation, that is the corresponding joint probability density function is given by a Fokker-Planck equation.
The model has been built phenomenologically, and the unclosed terms in the exact Lagrangian equations have been replaced 
by return to equilibrium and fluctuating terms, as in statistical mechanics when considering fluctuation-dissipation relations \citep{marconi2008fluctuation}.
More importantly, the model has been constructed in order for the Lagrangian pdf model to agree with all of the closed terms in the macroscopic Reynolds-average equations derived by \cite{Fox2014}. In addition, it provides closures for key unclosed terms, such as $\langle \mathbf{U}_s \rangle$, which play an important role in moderately dense fluid--particle flows such as CIT \citep{Fox2015}.  

In a companion paper, the Lagrangian pdf model developed in this work is applied to statistically homogeneous flows of increasing difficulty, namely, (i) particle-laden homogeneous isotropic turbulence \citep{fevrier2005partitioning,sundaram1999numerical}, (ii) homogeneous sheared turbulence \citep{elghobashi1994predicting}, and (iii) CIT \citep{Fox2015}. In the first two cases, the mean fluid- and particle-phase velocities are null, and hence the production of fluid-phase turbulence by fluid drag is absent. These cases are useful for validating the coupling terms in the Lagrangian pdf model for the exchange of turbulent kinetic energy between the two phases, and their dependence on the mass loading. In contrast, case (iii) provides a difficult validation test of the model for $\langle \mathbf{U}_s \rangle$, which determines the mean slip velocity between the two phases (see \eqref{eq:feq_Up}), and of the exchange/dissipation models, which determine the relative contributions of correlated $k_p$ and uncorrelated $\Theta_p$ turbulent kinetic energy.  

In future work, the spatial transport terms in the Lagrangian pdf model will be validated against particle-laden turbulent channel flow data with significant mass loading, such as in \cite{capecelatro2016strongly}.

\section*{Acknowledgments}

ROF was partially supported by grants from the U.S.\ National Science Foundation (CBET-1437865 and ACI-1440443).

\appendix

\section{Simulation of a Gaussian vector: the Choleski decomposition} \label{app:choleski}
Let $\mb{X}=(X_1,\dots,X_d)$ be a Gaussian vector defined by a zero
mean and a covariance matrix $C_{ij}=\lra{X_i X_j}$. For all positive
symmetric matrix (such as $C_{ij}$), there exists a (lower or upper)
triangular matrix $P_{ij}$ which satisfies
\begin{equation} \notag
\mb{C}=\mb{P}\mb{P}^t \,\Longrightarrow\, C_{ij} = \sum _{k=1}^{d}
P_{ik}P_{jk}.
\end{equation}
$\mb{P}$ is given by the Choleski algorithm (here for the lower
triangular matrix):
\begin{equation}
\begin{split}
& P_{i1} = \frac{C_{i1}}{\sqrt{C_{11}}},
           \quad 1 \leqslant i \leqslant d \notag \\
& P_{ii} = \left(C_{ii}-\sum _{j=1}^{i-1}P_{ij}\right)^{1/2},
           \quad 1 < i \leqslant d \\
& P_{ij} = \frac{1}{P_{jj}}
           \left(C_{ij}-\sum _{k=1}^{j-1}P_{ik}P_{jk}\right),
           \quad 1 < j < i \leqslant d \notag \\
& P_{ij} = 0, \quad i < j \leqslant d \notag .
\end{split}
\end{equation}
Let $\mb{G}=(G_1,\dots,G_d)$ be a vector composed of independent
$\mc{N}(0,1)$ Gaussian random variables, then it can be shown that the
vector $\mb{Y}=\mb{P}\mb{G}$ is a Gaussian vector of zero mean and
whose covariance matrix is $\mb{C}=\mb{P}\mb{P}^t$. Therefore,
$\mb{X}$ and $\mb{Y}$ are identical, that is
\begin{equation} \label{eq:simu}
\mb{X}=\mb{P}\mb{G}\,\Longrightarrow\, X_i
      = \sum _{k=1}^{d} P_{ik}G_{k}.
\end{equation}

\bibliographystyle{jfm}
\bibliography{paper}

\end{document}